\documentclass[12pt]{article}

\usepackage[T2A]{fontenc}
\usepackage{graphicx}
\usepackage{amsmath}
\usepackage{amssymb}
\usepackage[english]{babel}

\usepackage{enumitem}

\usepackage{caption}

\DeclareCaptionLabelSeparator{dot}{. }   

\begin{document}


\numberwithin{equation}{section}


\title{Exact solutions of a nonlinear diffusion equation \\ on polynomial invariant subspace of maximal dimension}
\author{S.\,R. Svirshchevskii \\
\ \\
    Keldysh Institute of Applied Mathematics \\
    Russian Academy of Sciences\\
    Moscow 125047, 
    Russia \\
    e-mail:
    {\tt svr@keldysh.ru}
}
\date{}
\maketitle

\begin{abstract}

The nonlinear diffusion equation $u_t = (u^{- 4/3}\, u_x)_x$ is reduced by the substitution $u = v^{- 3/4}$ to an equation with quadratic nonlinearities possessing a polynomial invariant linear subspace of the maximal possible dimension equal to five.
The dynamics of the solutions on this 
subspace is described by a fifth-order nonlinear dynamical system (V.A.~Galaktionov).

We found that, on differentiation,
this system reduces to a single linear equation of the second order, which is a special case of the Lam\'e equation,
and that the general solution of this linear equation is expressed in terms of the Weierstrass $\wp$-function and its derivative.
As a result, all exact solutions $v(x,t)$ on a five-dimensional polynomial invariant subspace,
as well as the corresponding solutions $u(x,t)$ of the original equation, are constructed explicitly.

Using invariance condition, two families of non-invariant solutions are singled out.
For one of these families, all types of solutions are considered in detail.
Some of them describe peculiar blow-up regimes, while others fade out in finite time.

\smallskip


\smallskip


\end{abstract}


\section{Introduction}

Nonlinear diffusion equations of the form
\begin{equation} \label{heat_m}
u_t = \nabla \cdot (D(u) \nabla u),
\end{equation}
where $u(x,t)>0$, $(x,t) \in R^N \times R$,
are found in many applications \cite{Cra1956}.
A rich variety of theoretical results are available for such equations, in particular, their group properties, various classes of exact solutions, etc., see, for example, \cite{IbrHandbook1994}.
Especially, equations with a power-law coefficient $D(u)$,
\begin{equation} \label{heat-m}
u_t = \nabla \cdot (u^m\, \nabla u)\,,
\end{equation}
 describing the processes of ``slow'' (for $m>0$) or ``fast'' (for $m<0$) diffusion, are widely used, see \cite{King1990, King1993} and the references therein.

The present paper is devoted to the construction of a class of exact solutions for the equation
\begin{equation} \label{heat}
u_t = (u^{- 4/3}\, u_x)_x \,,
\end{equation}
which was discussed by V.A.~Galaktionov in \cite{Gal}.
By substituting
\begin{equation} \label{u v}
u = v^{- 3/4},
\end{equation}
equation (\ref{heat}) is reduced to a quadratic form
\begin{equation} \label{Gal_polyn5_eq}
v_t = v v_{2} - \frac{3}{4}\,v_1^2 \equiv F[v]\,, \ \ v_i = \frac{\partial^i v}{\partial x^i}\,. 
\end{equation}
It was established in \cite{Gal} that the operator $F$
possesses the polynomial invariant subspace
\begin{equation} \label{polyn5}
  W_5 = \mathcal{L}\{1,x,\dots,x^4\}
\end{equation}
(a linear span of functions $1,x,\dots,x^4$),
that is, $F[W_5] \subseteq W_5$,
of the maximal possible dimension  equal to five
(recall that the dimension of a linear subspace invariant with respect to a nonlinear ordinary differential operator of the order $k$ cannot exceed $2 k + 1$;
see \cite{Svr} and \cite{GalSvr}, Ch.~2).
As a result, equation (\ref{Gal_polyn5_eq}) has solutions of the form
\begin{equation} \label{Gal_polyn5_sol}
v = C_0(t) + C_1(t) x + \frac{1}{2}\,C_2(t) x^2 + \frac{1}{6}\,C_3(t) x^3 + \frac{1}{24}\,C_4(t) x^4
\end{equation}
with coefficients $C_i(t)$, $i = \overline{\,0,4}$\,,
satisfying the dynamical system
 \begin{equation} \label{Gal_polyn5_DS}
\begin{aligned}
\dot C_0 &= C_0 \, C_2 - \frac{3}{4} \, C_1^2 = b_0,
\\
\dot C_1 &= C_0 \, C_3 - \frac{1}{2} \, C_1 C_2 = b_1,
\\
\dot C_2 &= C_0 \, C_4 + \frac{1}{2} \, (C_1 C_3 - C_2^2) = b_2,
\\
\dot C_3 &= \frac{3}{2} \, C_1 C_4 - \frac{1}{2} \, C_2 \, C_3 = b_3,
\\
\dot C_4 &= C_2 \, C_4 - \frac{1}{2} \, C_3^2 = b_4,
\end{aligned}
 \end{equation}
where the notation $b_0,\dots,b_4$ is introduced for the right-hand sides of the system.

The main purpose of the paper is to find
all solutions of  system (\ref{Gal_polyn5_DS}) and, as a consequence, all solutions of the form (\ref{Gal_polyn5_sol}) of equation (\ref{Gal_polyn5_eq}) and the corresponding solutions of equation (\ref{heat}).

Note that equation (\ref{heat}) possesses a five-dimensional Lie algebra of point symmetries \cite{Ovs1959eng}, which is maximal for equations of the form
\begin{equation*}
  u_t = (f(u) u_x)_x\,, \ \ f(u) \neq const\,.
\end{equation*}
The symmetry algebra of equation (\ref{Gal_polyn5_eq}) is also five-dimensional, and some of its solutions of the form (\ref{Gal_polyn5_sol}) turn out to be invariant with respect to some one-dimensional subalgebra 
and can be found by means of group analysis \cite{Ovs1978eng, Olv1989eng}.
Therefore, it is of particular interest to construct solutions of the form (\ref{Gal_polyn5_sol}) which are not invariant.

This paper extends
the study \cite{King1993} where a class of radially symmetric solutions of equation (\ref{heat-m}) of the form $u = v^{1/m}$ with a function $v$ polynomial in poweres of $r = |x|$
is described and, in particular, a number of new non-invariant solutions are constructed.

\smallskip

The content of the paper and the main ideas are as follows.

\vspace{-0.3cm}
\begin{itemize}[leftmargin=*]
\itemsep0em
\item 
In Section \ref{InvNoninv}, a simple criterion of invariance (non-invariance) for  solutions (\ref{Gal_polyn5_sol}) is obtained.
Using this criterion, all the obtained solutions are classified as invariant or non-invariant.


\smallskip
\item 
To construct solutions to system (\ref{Gal_polyn5_DS}), one can apply classical methods of reducing its order using the symmetry algebra or first integrals (the latter are given in Section~\ref{first_int}).

However, we use a simpler method based on the following remarkable property: on differentiation,
system (\ref{Gal_polyn5_DS}) is reduced to a single linear second-order equation
\begin{equation} \label{1_eqn_y}
\ddot{y}  = 2\, Q(t)\, y\,,
\end{equation}
where $Q(t)$ is a solution of the equation
\begin{equation}  \label{1_eqn_Q}
\dot{Q}^2 = 4 Q^3 - \tilde{g}_3\,, \ \
\tilde{g}_3 = const.
\end{equation}
All functions  $C_i(t)$, $i = \overline{\,0,4}$\,, must satisfy equation (\ref{1_eqn_y}), and the corresponding solution (\ref{Gal_polyn5_eq}) turns out to be non-invariant only if
$\tilde{g}_3 \neq 0$ (Sections~\ref{InvNoninv} and \ref{Q=P}.)


\smallskip
\item 
In the case $\tilde{g}_3 \neq 0$, the function $Q(t)$ is expressed in terms of the Weierstrass $\wp$-function, $Q(t) = \wp(t;0,\tilde{g}_3)$
(for $Q(t)=const$, system (\ref{Gal_polyn5_DS}) has only the trivial solution),
and equation (\ref{1_eqn_y}) is
a particular case of the Lam\'{e} equation \cite{Akh1990}.
In Section~\ref{lame_sol}, it is shown that in this case the general solution of equation (\ref{1_eqn_y}) is represented in the form
\begin{equation*}
y = \alpha\, \frac{1}{P(t)} +
\beta\, \frac{\dot{P}(t)}{P(t)}\,,
\end{equation*}
where $P(t) = \wp(t;0,g_3)$,
$g_3 = - \frac{1}{27}\,\tilde{g}_3$, and
$\alpha$, $\beta$ are arbitrary constants.
Note that this formula is much simpler than the well-known representation of the solution of the Lam\'{e} equation in terms of the Weierstrass $\sigma$-function \cite{Akh1990}.


\smallskip
\item 
It follows that the corresponding solutions of equation (\ref{Gal_polyn5_eq}) (in the case $\tilde{g}_3 \neq 0$) have the form
\begin{equation*}
v = A(x)\, \frac{1}{P(t)} + B(x)\, \frac{\dot{P}(t)}{P(t)}\,\,,
\end{equation*}
where $A(x)$ and $B(x)$ are fourth degree polynomials in $x$.
Having found them, we obtain two families of non-invariant solutions of equation (\ref{Gal_polyn5_eq}).
The first family (up to translations and dilations) is given by the formula
\begin{equation} \label{1_sol-1}
v = \frac{1}{4}\,\frac{1}{P(t)} \Bigl(S_1 + S_2 x^4\Bigr)
- \frac{1}{2}\, \dfrac{\dot{P}(t)}{P(t)}\, x^2
\,,
\end{equation}
where
$S_1$ and $S_2$ are arbitrary constants such that
\begin{equation*}
S_1 S_2 = - g_3 \neq 0\,.
\end{equation*}
The second family contains all powers of $x$ from zero to
four.
It is reduced to (\ref{1_sol-1}) by some transformations (generally speaking, complex) involving the
inversion transformation
 $ x \rightarrow 1/x$\,, \
 $ v \rightarrow v/x^{\,4}$
(Section~\ref{lame_sol}).


\smallskip
\item 
Note that solutions (\ref{1_sol-1}) are actually solutions on the three-dimensional invariant subspace $\mathcal{L}\{1,x^2,x^4\} \subseteq W_5$.
Solutions on this subspace are described by formulas (\ref{Gal_polyn5_sol}), (\ref{Gal_polyn5_DS}) with $C_1 = C_3 = 0$.
In this case, system (\ref{Gal_polyn5_DS}) is simplified and easily integrated.

In \cite{King1993}, a more general case of radially symmetric solutions of equation (\ref{heat-m}) of the form $u = v^{1/m}$,
$v = a_0(t) + a_1(t)\, r^2 + a_2(t)\, r^4$, where $r = |x|$,
was considered for the case $m = - \frac{4}{N+2}$
and arbitrary dimension $N$.
The existence of such solutions was proved in \cite{Gal}.
Solutions of the corresponding dynamical system for the coefficients $a_0(t)$, $a_1(t)$, $a_2(t)$ are expressed in \cite{King1993} in quadratures.

Below, in Section~\ref{King2}, an ``explicit'' representation of solutions of this dynamical system is given, and the corresponding solutions $v(x,t)$ are reduced to a form similar to (\ref{1_sol-1}), generalizing it to the case of  arbitrary $N$.
Conditions for invariance and non-invariance of solutions are indicated; the case $N=1$ corresponds to  non-invariant solution (\ref{1_sol-1}).


\smallskip
\item 
In Section~\ref{Ap_noninv-sol-1}, all types of solutions of the form (\ref{1_sol-1}) are studied in detail.
It is shown that some of them
describe peculiar blow-up regimes, while others fade out in  finite time.


\smallskip
\item 
Section~\ref{King1993} discusses another remarkable King's example, 
 \cite{King1993}, concerning equation (\ref{heat-m}) with $N=1$, $m = - \frac{3}{2}$\,.
The substitution $u = v^{1/m}$ reduces this equation to a quadratic form,
for which solutions are constructed on a four-dimensional polynomial invariant subspace.
The set of obtained solutions contains non-invariant solutions.
We represent these solutions in explicit form, expressing them in terms of the Weierstrass functions $\wp$ and $\zeta$,
and establish their non-invariance.


\smallskip
\item 
Along with non-invariant solutions, we give the corresponding invariant solutions in all cases. They are, of course, known and are presented for completeness.

\end{itemize}


\newpage

\section{Invariance and non-invariance conditions\\ for solutions (\ref{Gal_polyn5_sol})}
\label{InvNoninv}

Equation (\ref{Gal_polyn5_eq}) possesses a five-dimensional algebra of point symmetries with the basis
 \begin{equation} \label{Gal_polyn5_X1-X5}
\begin{gathered}
X_1 = \frac{\partial}{\partial t} \,, \ \ X_2 = t \frac{\partial}{\partial t} - v \frac{\partial}{\partial v} \,, \ \ X_3 = \frac{1}{2} \, x^2 \frac{\partial}{\partial x} + 2 x v \frac{\partial}{\partial v} \,,
\\[4pt]
X_4 = x \frac{\partial}{\partial x} + 2 v \frac{\partial}{\partial v} \,, \ \ X_5 = \frac{\partial}{\partial x} \,. \end{gathered}
 \end{equation}
The general form of the operator admitted by equation (\ref{Gal_polyn5_eq}) is as follows:
\begin{equation*}
 X = \alpha_1 X_1 + \dots + \alpha_5 X_5 =
 \xi(t) \frac{\partial}{\partial t} + \eta(x) \frac{\partial}{\partial x} + \sigma(x) v \frac{\partial}{\partial v}\,,
\end{equation*}
where
\begin{equation*}
\xi(t) = \alpha_1 + \alpha_2 t\,, \ \ \
 \eta(x) = \frac{1}{2}\,\alpha_3 x^2 + \alpha_4 x + \alpha_5 \,,
\ \
\sigma(x) = 2 \alpha_4 - \alpha_2 + 2 \alpha_3 x \,,
\end{equation*}
$\alpha_1, \dots, \alpha_5$
 are arbitrary constants that are not equal to zero at the same time.
%
A solution $v = \varphi(x,t)$ of equation (\ref{Gal_polyn5_eq}) is invariant with respect to the operator $X$ if and only if this solution satisfies the condition
$\sigma v = \xi v_t + \eta v_1$
or
\begin{equation}\label{inv_cond_gen}
\sigma v = \xi \Bigl(v v_2 - \frac{3}{4} \, v_1^2\Bigr) + \eta v_1\,.
\end{equation}
For solutions of the form (\ref{Gal_polyn5_sol}), condition (\ref{inv_cond_gen})
(after ``splitting'' in $x$) is reduced to the system
\begin{equation}\label{inv_cond}
\begin{aligned}
\xi b_0 &=  (2 \alpha_4 - \alpha_2) C_0 - \alpha_5 C_1, 
\\
\xi b_1 &= 2 \alpha_3 C_0 + (\alpha_4 - \alpha_2) C_1 - \alpha_5 C_2, 
\\
\xi b_2 &= 3 \alpha_3 C_1 - \alpha_2 C_2 - \alpha_5 C_3, 
\\
\xi b_3 &= 3 \alpha_3 C_2 - (\alpha_4 + \alpha_2) C_3 - \alpha_5 C_4, 
\\
\xi b_4 &= 2 \alpha_3 C_3 - (2 \alpha_4 + \alpha_2) C_4, 
\end{aligned}
\end{equation}
which must be fulfilled identically in $t$.

System (\ref{inv_cond}) is a system of linear equations with respect to the coefficients $\xi$, $\alpha_2$, $\alpha_3$, $\alpha_4$, $\alpha_5$, and the determinant of its matrix is equal to
\begin{equation}\label{delta0}
\begin{aligned}
\Delta &= \frac{1}{4}\, (32 C_0^3 C_4^3 - 96 C_0^2 C_1 C_3 C_4^2 - 96 C_0^2 C_2^2 C_4^2 + 144 C_0^2 C_2 C_3^2 C_4 
\\[2pt]
& - 36 C_0^2 C_3^4 + 216 C_0 C_1^2 C_2 C_4^2 - 12 C_0 C_1^2 C_3^2 C_4 - 240 C_0 C_1 C_2^2 C_3 C_4 
\\[4pt]
& + 72 C_0 C_1 C_2 C_3^3 + 72 C_0 C_2^4 C_4 - 24 C_0 C_2^3 C_3^2 - 81 C_1^4 C_4^2 
\\[4pt]
&+ 108 C_1^3 C_2 C_3 C_4 - 32 C_1^3 C_3^3 - 36 C_1^2 C_2^3 C_4 + 12 C_1^2 C_2^2 C_3^2).
\end{aligned}
\end{equation}
A direct verification shows that
$\dot{\Delta}\big|_{(\ref{Gal_polyn5_DS})} = 0$\,,
that is, $\Delta$ is the first integral of system (\ref{Gal_polyn5_DS}), 
$\Delta = const$ on every solution.

In the case $\Delta \neq 0$, system (\ref{inv_cond}) has only
the trivial solution
$\alpha_1 = \dots = \alpha_5 = 0$, therefore,
solution (\ref{Gal_polyn5_sol}) is not invariant.
If $\Delta = 0$, then
system (\ref{inv_cond}) can have a non-trivial constant solution $\alpha_1$, $\dots$, $\alpha_5$, and then
solution (\ref{Gal_polyn5_sol}) will be invariant.
Below, in Section~\ref{inv_sol}, it is shown that all solutions of the form (\ref{Gal_polyn5_sol}) corresponding to this case are invariant.
Thus, the following criterion for the invariance or non-invariance of solution (\ref{Gal_polyn5_sol}) holds.

\smallskip
\textbf{Proposition.} Solution (\ref{Gal_polyn5_sol}) is \textit{invariant} if $\Delta = 0$\,, and \textit{non-invariant} if $\Delta \neq 0$\,.

\section{All solutions of system (\ref{Gal_polyn5_DS}) and the corresponding \\ solutions (\ref{Gal_polyn5_sol})
of equation (\ref{Gal_polyn5_eq})}
\label{all_sol}

\subsection{Differential consequence of system (\ref{Gal_polyn5_DS})
and first integrals}
\label{first_int}

Differentiating equations (\ref{Gal_polyn5_DS}),
we obtain
\begin{equation} \label{DSt}
\ddot{C_i}  = \dot{b}_i = 2\, Q\, C_i\,, \ \ \ i=\overline{0,4}\,,
\end{equation}
where
\begin{equation}  \label{q}
2\, Q = C_0 C_4  - C_1 C_3  + \frac{1}{2}\,C_2^2
\end{equation}
(the coefficient ``2'' is introduced to simplify further calculations).

\smallskip
It follows that for any $i,j=\overline{0,4}$ it is fulfilled
\begin{equation*}
\ddot{C_i}\, C_j - C_i\,\ddot{C_j} = 0
\end{equation*}
or
\begin{equation*}
\frac{d}{d\,t} \bigl(\dot{C_i}\, C_j - C_i\,\dot{C_j}\bigr) =
\frac{d}{d\,t} \bigl(b_i\,C_j - C_i\,b_j\bigr) = 0\,.
\end{equation*}
Therefore, system (\ref{Gal_polyn5_DS}) possesses the first integrals
\begin{equation*}
C_j\,b_i - C_i\,b_j\,, \ \ \
i \neq j\,.
\end{equation*}
In particular, one can take the following integrals:
\begin{equation*}
\begin{aligned}
p_1 &= C_0 b_1 - C_1 b_0 = \frac{1}{4}\, (4 C_0^2 C_3 - 6 C_0 C_1 C_2 + 3 C_1^3)\,,
\\[4pt]
p_2 &= C_0 b_2 - C_2 b_0 = \frac{1}{4}\, (4 C_0^2 C_4 + 2 C_0 C_1 C_3 - 6 C_0 C_2^2 + 3 C_1^2 C_2)\,,
\\[3pt]
p_3 &= C_0 b_3 - C_3 b_0 = \frac{3}{4}\,(2 C_0 C_1 C_4 - 2 C_0 C_2 C_3 + C_1^2 C_3)\,,
\\[3pt]
p_4 & = C_0 b_4 - C_4 b_0 = \frac{1}{4}\, ( - 2 C_0 C_3^2 + 3 C_1^2 C_4)\,.
\end{aligned}
\end{equation*}
From this system, one can express the functions $C_1(t),\dots,C_4(t)$ in terms of $C_0(t)$ and arbitrary constants $p_1,\dots,p_4$, then, solving the equation for $C_0(t)$, find all solutions to system (\ref{Gal_polyn5_DS})
and eventually get all solutions of the form
(\ref{Gal_polyn5_sol}) for equation (\ref{Gal_polyn5_eq}).

However, we use a different, shorter way.

\subsection{Equation for $Q(t)$ and its solutions}
\label{Q=P}

Differentiating (\ref{q}) by virtue of system (\ref{Gal_polyn5_DS}), yields
\begin{equation*}
\dot{Q} = \frac{1}{8}\,\left(12\, C_0 C_2 C_4 - 9\, C_1^2 C_4 - 6\,C_0 C_3^2 + 6\,C_1 C_2 C_3 - 2\, C_2^3\right).
\end{equation*}
It is directly verified that
\begin{equation}  \label{eqn_Q}
\dot{Q}^2 = 4 Q^3 - \tilde{g}_3\,, \ \
\tilde{g}_3 = \frac{1}{16}\,\Delta\,,
\end{equation}
where $\Delta$ is defined by (\ref{delta0}).

For each solution $Q(t)$ of equation (\ref{eqn_Q}), one can solve system
(\ref{DSt}) with respect to the functions $C_i(t)$ and, as a result, find all solutions (\ref{Gal_polyn5_sol}) of equation (\ref{Gal_polyn5_eq}).

Thus, the following cases arise (we write formulas up to translations in t):

\smallskip
1) $\Delta \neq 0$\,.
Then the unique
a non-constant solution to equation (\ref{eqn_Q})
is expressed in terms of the Weierstrass $\wp$-function:
\[Q(t) = \wp(t;0,\tilde{g}_3)\,.\]
In this case, the corresponding solutions (\ref{Gal_polyn5_sol}) to equation (\ref{Gal_polyn5_eq}) are non-invariant (see Proposition in Section~\ref{InvNoninv}).

Equation (\ref{eqn_Q}) also possesses
constant solution
$Q(t) = \Bigl(\frac{1}{4}\,\tilde{g}_3\Bigr)^{1/3} \neq 0$.
It is easily seen that only the zero solution $C_0 = \dots = C_4 = 0$ of system (\ref{Gal_polyn5_DS}) corresponds to this case, in contradiction with assumption $\Delta \neq 0$ 
(we omit the calculations).

\smallskip
2) $\Delta = 0$\,.
Then the only solutions of equation (\ref{eqn_Q}) are the following:
\begin{equation*}
 Q(t) \equiv 0 \ \ \, \text{and} \ \ \, Q(t) = t^{-2}\,.
\end{equation*}
In these cases, all the corresponding solutions (\ref{Gal_polyn5_sol}) of equation (\ref{Gal_polyn5_eq}) turn out to be invariant (Section~\ref{inv_sol}).

\subsection{Case $\Delta \neq 0$, $Q(t) = \wp(t;0,\tilde{g}_3)$ (non-invariant solutions)}
\label{lame_sol}

In this case, each equation of system (\ref{DSt}) is
a linear Lam\'e equation of the form
\begin{equation} \label{eqn_y}
\ddot{y}  = 2\, Q(t)\, y\,,
\end{equation}
where 
$Q(t) = \wp(t;0,\tilde{g}_3)$, $\tilde{g}_3 \neq 0$.
This is a special case of the general Lam\'e equation, the solutions of which are known to be expressed in terms of the Weierstrass $\sigma$-function, see \cite{Akh1990}.
For equation (1), we obtain a simpler formula for the general solution.

Equation (\ref{eqn_y}) will be considered on the interval $(0,T)$, where $T$ is the period of the function $Q(t)$.
Along with the function $Q(t)$, consider the function $P(t) = \wp(t;0,g_3)$ with $g_3 = - \frac{1}{27}\,\tilde{g}_3$, which is a non-constant solution to equation
\begin{equation}  \label{eqn_P}
\dot{P}^2 = 4 P^3 - g_3\,, \ \
g_3 = - \frac{1}{432}\,\Delta\,.
\end{equation}
Note that the period $\tau$ of $P(t)$ is equal to $T$ if $g_3 > 0$ and $3\,T$ if $g_3 < 0$.

We show that the general solution of equation (\ref{eqn_y}) is expressed in terms of the function $P(t)$ and its derivative $\dot{P}(t)$.

\smallskip
\textbf{Lemma.}
The following formula holds:
\begin{equation}  \label{QP}
Q(t) = P(t)- g_3 P^{-2}(t), \ \ t \in (0,\,T).
\end{equation}

\medskip
\noindent
$\lhd$ \
Denoting $P(t)- g_3 P^{-2}(t) \equiv \widetilde{Q}(t)$,
we obtain
\begin{equation*}
\left(\dot{\widetilde{Q}}\right)^2 - 4 \left(\widetilde{Q}\right)^3 + \tilde{g}_3
\equiv \Bigl(\dot{P}^2 - 4 P^3 + g_3\Bigr)
\Bigl( 1 + 2 g_3 P^{-3} \Bigr)^2,
\end{equation*}
so $\widetilde{Q}(t)$ is a (non-constant) solution to equation (\ref{eqn_Q}).
Hence, $\widetilde{Q}(t) \equiv Q(t + C)$, where $C = const$.
For $t \rightarrow 0$ we have $\widetilde{Q}(t) \rightarrow +\infty$ and $Q(t) \rightarrow +\infty$.
Therefore, $C = 0$ and $\widetilde{Q}(t) \equiv Q(t)$.
\hfill $\blacktriangleright$

\medskip
\textbf{Remark~1.}
If $g_3 > 0$, then for a given function $Q(t)$ there exist a \textit{unique} function $P(t)$ satisfying (\ref{QP}).

If $g_3 < 0$, then there are \textit{three} such functions:
along with $P(t)$ in (\ref{QP}), one can 
also use the functions $P(t + T)$ and $P(t + 2 T)$,
or, equivalently, use the same function $P(t)$,
taken on intervals $(T,\,2 T)$ and $(2 T,\,3 T)$
(we will apply this remark in Section~\ref{Ap_noninv-sol-1}).

\medskip

\textbf{Remark~2.} \smallskip
From (\ref{QP}), we obtain the following identity for the $\wp$\,-function:
\begin{equation}  \label{SS_ident}
\wp(t; 0, g_3) - g_3\,\wp^{-2}(t; 0, g_3) =
\wp(t; 0, - 27 g_3),
\end{equation}
which is valid on the complex plane.
Transforming the right-hand side according to the well-known formula
$\wp(t; 0, \varepsilon^{6} g_3) = \varepsilon^{2} \wp(\varepsilon t; 0, g_3)$
for
$\varepsilon = i \sqrt{3}$,
we obtain the identity
\begin{equation*}
\wp(t; 0, g_3) - g_3\,\wp^{-2}(t; 0, g_3) =
- 3\, \wp(i \sqrt{3}\, t; 0, g_3),
\end{equation*}
proved in \cite{GomFerNov2021} in another way.

\medskip
Given the representation (\ref{QP}), it can be verified that the functions
$1 / P(t)$ and $\dot{P}(t) / P(t)$ form a fundamental system of solutions to equation (\ref{eqn_y}).
This implies the following result.

\medskip
\textbf{Theorem.}
The general solution of the Lam\'e equation (\ref{eqn_y}) on the interval $(0,\,T)$ has the form
\begin{equation*}
y = \alpha\, \frac{1}{P} +
\beta\, \frac{\dot{P}}{P}\,,
\end{equation*}
where $\alpha$ and $\beta$ are arbitrary constants.

\medskip

Returning to system (\ref{DSt}), we find its general solution
\begin{equation} \label{Ci}
C_i = \alpha_i\, \frac{1}{P} +
\beta_i\, \frac{\dot{P}}{P}\,,  \ \ \ i=\overline{0,4}\,,
\end{equation}
with arbitrary constants $\alpha_i$, $\beta_i$\,.
Substituting these expressions in (\ref{Gal_polyn5_DS}) and splitting
the obtained identities in powers of $\dot{P}$ leads to a system of algebraic equations, from which
it is not difficult to find the coefficients $\alpha_i$, $\beta_i$.

The calculations, however, can be simplified by proceeding
as described below.

The solution (\ref{Gal_polyn5_sol}) of equation (\ref{Gal_polyn5_eq}) corresponding to (\ref{Ci}) has the form
\begin{equation} \label{v_AB4}
v = A(x)\, \frac{1}{P} +
B(x)\, \frac{\dot{P}}{P}\,\,,
\end{equation}
where A and B are functions of $x$
(polynomials of the fourth degree, but this does not need to be assumed in advance: they will turn out to be just that).

Substitution of (\ref{v_AB4}) into (\ref{Gal_polyn5_eq}) taking into account equation (\ref{eqn_P}) and its corollary $\ddot{P} = 6 P^2$ yields
\begin{equation*}
\begin{gathered}
  - A \dot{P} + \frac{1}{2}\,B \left(\dot{P}^2 + 3 g_3 \right) =
   A A^{\prime\prime} - \frac{3}{4}\, {A^{\prime}}^{\,2}
+
\\[4pt]
+  \left( A B^{\prime\prime} + A^{\prime\prime} B - \frac{3}{2}\, A^{\prime} B^{\prime} \right) \dot{P}
+ \left( B B^{\prime\prime} - \frac{3}{4}\, {B^{\prime}}^{\,2} \right) \dot{P}^2.
\end{gathered}
\end{equation*}
Equating the coefficients at the powers of $\dot{P}$ on the left- and right-hand sides, we obtain the system
\begin{equation} \label{system_A-B}
  \begin{gathered}
B B^{\prime\prime} - \frac{3}{4}\, {B^{\prime}}^{\,2} = \frac{1}{2}\,B,
\\[2pt]
A B^{\prime\prime} + B A^{\prime\prime} - \frac{3}{2}\, A^{\prime} B^{\prime}  = - A,
\\[2pt]
A A^{\prime\prime} - \frac{3}{4}\, {A^{\prime}}^{\,2} = \frac{3}{2}\,g_3 B.
  \end{gathered}
\end{equation}

\smallskip
To bring solutions to the simplest possible form, we will use the following symmetries of equation (\ref{Gal_polyn5_eq}).

\smallskip
{a)}
The dilation
\begin{equation} \label{dilation1}
  t = \theta\, \tilde{t}\,, \ \ v = \theta^{-1} \tilde{v} \ \ (\theta > 0) 
\end{equation}
reduces solution (\ref{v_AB4}) to the form
\begin{equation*}
\tilde{v} = \theta^3 A(x)\, \frac{1}{\widetilde{P}} +
B(x)\, \frac{\dot{\widetilde{P}}}{\widetilde{P}}\,,
\end{equation*}
where
$\widetilde{P}\left(\tilde{t}\,\right) = \theta^{2} P(t)$
satisfies the equation
\begin{equation*}
\bigl(\dot{\widetilde{P}}\bigr)^2 = 4 (\widetilde{P})^3 - \tilde{g}_3\,, \ \ \tilde{g}_3 = \theta^6 g_3\,.
\end{equation*}
For $g_3 \neq 0$, choosing $\theta = |g_3|^{-1/6}$, we get $|\tilde{g}_3| = 1$.
Therefore, in what follows we can assume that $g_3 = \pm 1$\,.
However, for ease of use, in formulas for non-invariant solutions we leave $g_3$ an arbitrary constant.

\smallskip
{b)}
The dilation
\begin{equation} \label{equiv2}
x = a\, \tilde{x}, \ \ v = a^2\, \tilde{v} \ \
(a > 0)
\end{equation}
reduces solution (\ref{v_AB4}) to the form
\begin{equation*}
\tilde{v} = \tilde{A}(\tilde{x})\, \frac{1}{P} +
\tilde{B}(\tilde{x})\,\frac{\dot{P}}{P}
\,,
\end{equation*}
where $\tilde{A}(\tilde{x}) = a^{-2} A(x)$\,,
$\tilde{B}(\tilde{x}) = a^{-2} B(x)$.
System (\ref{system_A-B}) also admits the corresponding dilation
$x = a \tilde{x}$, $A = a^2 \tilde{A}$,
$B = a^2 \tilde{B}$\,.

\smallskip
Up to translations in $x$ and dilations (\ref{equiv2}),
the first equation of system (\ref{system_A-B}) has
 the following solutions:
\begin{equation*}
B = 0, \ \ B = - \frac{1}{2}\, x^2 \ \ \text{and} \ \ B = \frac{1}{8}\,\delta \left(x^2 + \delta \right)^2, \ \delta = \pm 1.
\end{equation*}
For each of these solutions, we find the function $A(x)$ that satisfies the two remaining equations of system (\ref{system_A-B})
and obtain the corresponding solutions (\ref{v_AB4}) of equation (\ref{Gal_polyn5_eq}).
We have three cases:

\smallskip
1) $B = 0$\,.
It follows
that $A = 0$ and the corresponding solution (\ref{v_AB4}) is trivial.

2) $B = - \dfrac{1}{2}\,x^2$\,.
The corresponding solutions (\ref{v_AB4}) have the form
\begin{equation} \label{sol-1} 
v = \frac{1}{4} \, \frac{1}{P} \, \Bigl(S_1 + S_2 x^4\Bigr)
- \frac{1}{2}\, \frac{\dot{P}}{P}\, x^2
\,,
\end{equation}
where $S_1$ and $S_2$ are arbitrary constants, satisfying the condition
\begin{equation} \label{S1*S2}
S_1 S_2 = - g_3 \neq 0\,.
\end{equation}
Dilation (\ref{equiv2}) with $a = |S_2|^{-1/2}$
reduces (\ref{sol-1}), (\ref{S1*S2}) to
\begin{equation} \label{sol-1_simple}
v = \dfrac{\delta}{4}\, \dfrac{1}{P} \left(x^4 - g_3 \right)
  - \dfrac{1}{2}\, \dfrac{\dot{P}}{P}\, x^2\,,
\ \ \delta = \pm1\,.
\end{equation}

3) $B = \dfrac{1}{8}\,\delta \left(x^2 + \delta \right)^2, \ \delta = \pm 1$.
The corresponding solutions (\ref{v_AB4}) have the form
\begin{equation} \label{sol-2}
v = \frac{1}{8}\,  \frac{1}{P} \Bigl(4\, S_1 \left(x^3 - \delta x\right)
+ S_2\left(x^4 - 6 \delta x^2 +1\right)\Bigr)
+ \frac{\delta}{8}\, \frac{\dot{P}}{P} \left(x^2 + \delta \right)^2
\,,
\end{equation}
where $S_1$ and $S_2$ are arbitrary constants, satisfying the condition
\begin{equation} \label{S1+S2}
\delta S_1^2 + S_2^2 = - g_3 \neq 0\,.
\end{equation}

Formulas (\ref{sol-1}), (\ref{S1*S2}) and (\ref{sol-2}), (\ref{S1+S2}) represent two families of non-invariant solutions of equation (\ref{Gal_polyn5_eq}).
The first family is related to King's non-invariant solutions,
which we discuss in the next section.
A complete study of this family is given in Section~\ref{Ap_noninv-sol-1}.

\smallskip
\textbf{Remark~3.}
Solutions (\ref{sol-1}) and (\ref{sol-2}) are not invariant under any subalgebras of the five-dimensional algebra of continuous symmetries (\ref{Gal_polyn5_X1-X5}) of equation (\ref{Gal_polyn5_eq}).
But they admit some discrete symmetries of this equation.
%
For example, any solution of the form (\ref{sol-1}) is invariant under the reflection
\begin{equation}  \label{refl-1}
  x \rightarrow - x.
\end{equation}
Solution (\ref{sol-1_simple}) for $g_3 = - 1$ also admits the
transformation
\begin{equation}  \label{inv-1}
  x \rightarrow \dfrac{1}{x}\,, \
  v \rightarrow \dfrac{v}{x^{\,4}}\,,
\end{equation}
and for $g_3 = 1$ -- the composition of
(\ref{inv-1})
 and reflection
\begin{equation}  \label{refl-2}
  t \rightarrow - t, \ \ v \rightarrow - v.
\end{equation}
Similarly, any solution of the form (\ref{sol-2}) for $\delta = - 1$ admits
transformation (\ref{inv-1}),
and for $\delta = 1$ -- the composition of (\ref{inv-1})
and reflection (\ref{refl-1}).

\medskip
\textbf{Remark~4.}
Note that solutions (\ref{sol-2}) and (\ref{sol-1}) are connected
by some transformations, possibly complex.
Indeed, consider two cases:

\smallskip
a) $\delta = - 1$.
Introducing constants $\tilde{S}_1$ and $\tilde{S}_2$ such that
$S_1 = \frac{1}{2}\,(\tilde{S}_2 - \tilde{S}_1)$,
$S_2 = \frac{1}{2}\,(\tilde{S}_2 + \tilde{S}_1)$, we rewrite solution (\ref{sol-2}) in the form
\begin{equation*}
v = \frac{1}{16}\,\frac{1}{P}\,
\Bigl(\tilde{S}_1 (x-1)^4 + \tilde{S}_2 (x+1)^4\Bigr)
- \frac{1}{8}\, \frac{\dot{P}}{P}\,(x^2 - 1)^2
\,,
\ \
\tilde{S}_1 \tilde{S}_2 = - g_3.
\end{equation*}
Applying transformations $x \rightarrow x + 1$,
 (\ref{inv-1}) and $x \rightarrow x - 1/2$, yields
\begin{equation*}
v = \dfrac{1}{4}\, \frac{1}{P}\,
\Bigl(\dfrac{1}{4}\,\tilde{S}_1 + 4 \tilde{S}_2\, x^4\Bigr)
- \dfrac{1}{2}\, \frac{\dot{P}}{P}\, x^2
\,.
\end{equation*}
Substituting
$\dfrac{1}{4}\,\tilde{S}_1 \rightarrow S_1$,
$4\, \tilde{S}_2 \rightarrow S_2$\,,
we obtain solution (\ref{sol-1}).

\smallskip
b) $\delta = 1$.
Then $g_3 < 0$ (see (\ref{S1+S2})).
The complex transformation
$x \rightarrow i x$, $v \rightarrow - v$,
$S_1 \rightarrow - i\, S_1$, $S_2 \rightarrow - S_2$
changes the sign of $\delta$ in formulas (\ref{sol-2}), 
(\ref{S1+S2})
and leads to the previous case.


\subsection{Case $\Delta = 0$ (invariant solutions)}
\label{inv_sol}

Consider two cases.
(We write down solutions up to translations in $x$ and $t$, dilations (\ref{dilation1}), (\ref{equiv2}) and reflections (\ref{refl-1}), (\ref{refl-2}).)

\smallskip
1) $Q(t) = 0$.
In this case, the general solution of system (\ref{DSt}) is given by the formulas $C_i = \alpha_i\, t + \beta_i$,  $i=\overline{0,4}$\,, with arbitrary constants  $\alpha_i$, $\beta_i$, and the corresponding solution (\ref{Gal_polyn5_sol}) to equation (\ref{Gal_polyn5_eq}) has the form
\begin{equation*}
v = A(x)\, t + B(x),
\end{equation*}
where $A(x)$ and $B(x)$
are some functions (polynomials of degree four).
The substitution into (\ref{Gal_polyn5_eq}) and splitting
in $t$ yields a system for $A(x)$ and $B(x)$.
Solving this system, we obtain
\begin{equation*}
\text{(a)} \ \,
v = 1, \ \, v = x^4
\ \ \ \text{and} \ \,
\ \text{(b)} \ \,
v = \dfrac{4}{3}\,(x - t), 
\ \,
v = \dfrac{4}{3}\,x^3(1-x\, t). 
\end{equation*}
Note that solutions in each pair are connected by
transformation (\ref{inv-1}).

\smallskip
2) $Q(t) = t^{-2}$.
In this case, $g_3 = 0$ and from equation (\ref{eqn_P}),
we find $P(t) = t^{-2}$.
As in Section~\ref{lame_sol},
the corresponding solution (\ref{Gal_polyn5_sol}) of equation (\ref{Gal_polyn5_eq})
is given by the formula (\ref{v_AB4}),
\begin{equation*}
v = A(x)\, t^2 - 2 B(x) \frac{1}{t}\,,
\end{equation*}
with $A(x)$ and $B(x)$ defined by system (\ref{system_A-B}).
Non-trivial solutions in this case also have the form (\ref{sol-1}) and (\ref{sol-2}), but with $g_3 = 0$. Having considered them, we get the following solutions:
\begin{center}
(a) \
$v = \dfrac{x^2}{t}$\,,
\ \
\text{(b)} \
$v = t^2 + \dfrac{x^2}{t}$\,,
\
$v = t^2 x^4 + \dfrac{x^2}{t}$\,,

\medskip

(c) \
$v = t^2 \left(x + 1 \right)^4
+ \dfrac{1}{4 t}\, (x^2 - 1)^2$\,,
\
(d) \
$v = - \dfrac{\delta}{4 t}\, (x^2 + \delta)^2\,, \ \delta = \pm 1$\,.
\end{center}
Here solutions (b) are related by transformation (\ref{inv-1}),
and solution (c) is transformed by a composition of (\ref{inv-1}) and translations
into solution (a).
Solution (d) for $\delta = -1$ is also transformed by a composition of
(\ref{inv-1}) and translations into solution (a), and case $\delta = 1$ is related with the previous one by the complex change $x \rightarrow i x$, $v \rightarrow - v$.

\smallskip
It is directly seen that all solutions given in Section~\ref{inv_sol} are invariant.


\section{King's solutions for equation (\ref{heat-m}) with $m = - \frac{4}{N+2}$}
\label{King2}

Following \cite{King1993}, we consider a radially symmetric version of equation (\ref{heat-m}):
\begin{equation} \label{K2_heat_n_m}
u_t = \frac{1}{r^{N-1}}\, (r^{N-1} u^m u_r)_r,
\end{equation}
where $N$ is the number of spatial variables,
$r = \left( \sum_{i=1}^{N} x_i^2 \right)^{1/2}$.
By substituting $u = v^{1/m}$, equation (\ref{K2_heat_n_m}) is reduced to a quadratic form
\begin{equation} \label{K2_quadr_n_m}
v_t = v v_{rr} + \frac{1}{m}\,v_r^2 + \frac{N-1}{r}\,v v_r\,. %
\end{equation}

Consider the case
\begin{equation} \label{K2_m}
  m = - \dfrac{4}{N+2}\,.
\end{equation}
Note that in this case, equation (\ref{K2_quadr_n_m}) admits the
transformation
\begin{equation}  \label{K2 inv}
   \bar{r} = \dfrac{1}{r}\,, \
   \bar{v} = \dfrac{v}{r^{\,4}}\,,
\end{equation}
which we use below when discussing a set of
invariant solutions.

In \cite{King1993} (pp. 41, 42) solutions of the form
\begin{equation} \label{K2_sol}
v = a_0(t) + a_1(t)\, r^2 + a_2(t)\, r^4
\end{equation}
with coefficients satisfying the system
 \begin{equation} \label{K2_DS_n_m}
\dot a_0 = 2 N a_0 a_1,
\ \ \
\dot a_1 = (N-2)a_1^2 + 4(N+2) a_0 a_2,
\ \ \
\dot a_2 = 2 N a_1 a_2
 \end{equation}
were constructed for equation (\ref{K2_quadr_n_m}), (\ref{K2_m}).
The existence of such solutions was proved in \cite{Gal}, Theorem 3.3, where the invariance of a three-dimensional subspace $\mathcal{L}\{1,r^2,r^4\}$ with respect to the operator
$F[v] = v v_{rr} + \frac{1}{m}\,v_r^2 + \frac{N-1}{r}\,v v_r$
is established under condition (\ref{K2_m}) (see also \cite{GalSvr}, Proposition 6.24).

In \cite{King1993}, system (\ref{K2_DS_n_m}) is solved in quadratures.
Below an ``explicit'' representation of its solutions is given, and the corresponding solution (\ref{K2_sol}) is reduced to a form similar to
(\ref{sol-1}), generalizing it to the case of arbitrary N.
Also, a condition for the non-invariance of the solution is formulated and both non-invariant and invariant solutions of the form (\ref{K2_sol}) are indicated.

\subsection{All solutions of system (\ref{K2_DS_n_m}) and corresponding solutions (\ref{K2_sol}) of equation (\ref{K2_quadr_n_m})}

System (\ref{K2_DS_n_m}) is easily integrated.
Omitting calculations, we formulate the results.

The general solution to system (\ref{K2_DS_n_m})
is represented as
\begin{equation*}
a_0 = \frac{S_1}{4 N}\, \frac{1}{y}\,, \ \
a_1 = - \frac{1}{2 N}\, \frac{\dot{y}}{y}\,,
\ \ a_2 = \frac{S_2}{4 N}\, \frac{1}{y}\,,
\end{equation*}
where $y(t) \neq 0$ is a solution of the equation
\begin{equation} \label{K2_eq_y_2_k}
y \ddot{y} + k\, (\dot{y})^2 = k\, \alpha, \ \ k = - \frac{N+2}{2 N}\,,
\ \  \alpha = S_1 S_2,
\end{equation}
$S_1$, $S_2$ are arbitrary constants.
Multiplying (\ref{K2_eq_y_2_k}) by $2 y^{2 k-1} \dot{y}$ and integrating,  yields
\begin{equation} \label{K2_eq_y}
(\dot{y})^2 = \alpha + \beta \, y^{\frac{N+2}{N}}
\,,
\end{equation}
where $\beta$ is an arbitrary constant.

The corresponding solution (\ref{K2_sol}) of equation (\ref{K2_quadr_n_m}) has the form
\begin{equation} \label{K2_v_y_gen}
v = \dfrac{1}{4 N} \dfrac{1}{y}
\left( S_1 + S_2 r^4 \right)
- \dfrac{1}{2 N} \dfrac{\dot{y}}{y}\,r^2  \,,
\ \ S_1 S_2 = \alpha
\,,
\end{equation}
where $S_1$, $S_2$, $\alpha$, $\beta$ are arbitrary constants, $y(t) \neq 0$ is a solution of equation (\ref{K2_eq_y_2_k}) (or equation (\ref{K2_eq_y})).

It is shown below that solution (\ref{K2_v_y_gen}) is non-invariant
iff the condition $\alpha \beta \neq 0$ holds.
In this case, for $N=1$, the solution $y(t)$ of equation (\ref{K2_eq_y}) is
expressed in terms of the Weierstrass $\wp$-function and is defined on a bounded 
interval of $t$, and solution (\ref{K2_v_y_gen}), up to dilatation and reflection, coincides with the above non-invariant solution (\ref{sol-1}).
If $N \geq 2$, then the solution $y(t)$ of equation (\ref{K2_eq_y}) is defined for any real $t$, and for $N = 2$ it is expressed in elementary functions (see \cite{King1993}).

\subsection{
Condition for non-invariance of solutions (\ref{K2_v_y_gen})}
\label{Ap_K2_noninv}

Setting $a_0 = C_0$, $a_1 = \frac{1}{2}\, C_2$,
$a_2 = \frac{1}{24}\, C_4$,
we represent solution (\ref{K2_sol}) in the form similar to (\ref{Gal_polyn5_sol}):
\begin{equation} \label{K2_C}
v = C_0(t) + \frac{1}{2}\,C_2(t) r^2 + \frac{1}{24}\,C_4(t) r^4.
\end{equation}
In this case, system (\ref{K2_DS_n_m}) takes the form
\begin{equation} \label{K2_DS}
\dot C_0 = N C_0 \, C_2, 
\ \ \
\dot C_2 = \frac{N-2}{2} \, C_2^2 + \frac{N+2}{3} \, C_0 \, C_4, 
\ \ \
\dot C_4 = N C_2 \, C_4 
 \end{equation}
and its general solution (see (\ref{K2_v_y_gen})) is written as follows:
\begin{equation} \label{K2_C0,1,2}
C_0 = \frac{S_1}{4 N}\,\frac{1}{y}\,, \ \ \
C_2 = - \frac{1}{N}\, \frac{\dot{y}}{y}\,, \ \ \
C_4 = \frac{6 S_2}{N}\, \frac{1}{y}\,.
\end{equation}

The algebra of point symmetries of equation (\ref{K2_quadr_n_m}) differs for different $N$.
For $N \geq 3$ this algebra is three-dimensional and has a basis
\begin{equation*}
\begin{gathered}
X_1 = \frac{\partial}{\partial t} \,, \ \ X_2 = t \frac{\partial}{\partial t} - v \frac{\partial}{\partial v} \,, \ \
X_4 = r \frac{\partial}{\partial r} + 2 v \frac{\partial}{\partial v}
\end{gathered}
 \end{equation*}
(the numbering of the operators corresponds to (\ref{Gal_polyn5_X1-X5})).
For $N = 2$, the algebra is four-dimensional: its basis, in addition to
operators $X_1$, $X_2$, $X_4$ includes operator
 \begin{equation*}
X_0 = r \ln r \frac{\partial}{\partial r} + 2 (\ln r +1) v \frac{\partial}{\partial v}\,.
 \end{equation*}
For $N = 1$, the algebra is five-dimensional and has basis (\ref{Gal_polyn5_X1-X5})
(this case was considered above
in Section~\ref{InvNoninv}).

Acting as in Section~\ref{InvNoninv}, from the
invariance condition for solution (\ref{K2_C}),
we obtain a system of linear homogeneous equations for the coefficients of a linear combination of the basis operators.
The determinant of the matrix of this system (taking into account (\ref{K2_C0,1,2}) and (\ref{K2_eq_y})) has the form
\begin{equation*}
\Delta = 2 (N+2) C_0 C_4 \left( C_2^2 - \frac{2}{3}\, C_0 C_4\right)
= 3 \frac{N+2}{N^4}\, \alpha \beta\,
y^{\frac{2}{N}-3} \ \, \text{for} \ \, N \geq 2
\end{equation*}
and
\begin{equation*}
\Delta = 18\, C_0 C_4 \left( C_2^2 - \frac{2}{3}\, C_0 C_4\right)^2
= 9 \alpha  \beta^{\,2} \ \, \text{for} \ \, N = 1.
\end{equation*}

In the case $\Delta \neq 0$, solution (\ref{K2_C}) is non-invariant and hence the condition $\alpha \beta \neq 0$ is sufficient for non-invariance of solution (\ref{K2_v_y_gen}).
Since all solutions obtained from (\ref{K2_v_y_gen}), (\ref{K2_eq_y}) for $\alpha \beta = 0$
are invariant (see Section~(\ref{Ap_K2_inv})), the following statement holds.

\medskip
\textbf{Proposition.}
Solution (\ref{K2_v_y_gen}) is \textit{invariant} if $\alpha \beta = 0$\,, and \textit{non-invariant} if $\alpha \beta \neq 0$\,.

\subsection{Invariant solutions of the form (\ref{K2_v_y_gen})}
\label{Ap_K2_inv}

Having solved equation (\ref{K2_eq_y}) with $\alpha\, \beta = 0$, we find the corresponding solutions (\ref{K2_v_y_gen})
(where $S \neq 0$, $A \neq 0$, $S_0$ are arbitrary constants):

\medskip
\noindent
1) ($\alpha = \beta = 0$) \
$v = S$, \ $v = S\, r^4$\,;

\medskip
\noindent
2) ($\alpha = 0$, $\beta \neq 0$)
a)
$v = S_0\, e^{4 A\, t} + A\,r^2$, \
$v = S_0\, e^{4 A\, t} r^4 + A\,r^2$, \
if $N = 2$\,;

\medskip
b)
$v = S_0\, |t|^{\frac{2 N}{2 - N}} + \frac{1}{2-N}\,\frac{r^2}{t}$\,, \
$v = S_0\, |t|^{\frac{2 N}{2 - N}}\, r^4 + \frac{1}{2-N}\,\frac{r^2}{t}$\,, \
if $N \neq 2$\,;

\medskip
\noindent
3) ($\alpha \neq 0$, $\beta = 0$) \
$v = \frac{1}{4 N S\, t} \left( r^2 - S \right)^2$\,.

\medskip
Note that solutions in pairs 1), 2\,a) and 2\,b)
are connected by transformation (\ref{K2 inv}), and
solution 3) is transformed by (\ref{K2 inv})
into a solution of the same type, but
with the replacement $S \to 1/S$.

It is seen that all these solutions are invariant.
They are all known. In particular, solutions 2\,b) correspond to the solution of the ``instantaneous source'' and the ``dipole'' solution of equation (\ref{K2_heat_n_m}),
see \cite{ZelKom1950}, \cite{Bar1952, BarZel1957, Hil1989},
as well as \cite{King1990, King1993} and the references therein.

\subsection{Remark on the second derivatives of 
$C_i(t)$}

By virtue of system (\ref{K2_DS}), for the second derivatives of the functions $C_i(t)$ we obtain
\begin{equation} \label{K2_DS'}
\ddot C_i = \mu\, \dot{C}_i + \nu \, C_i, \ \ \ i = 0,2,4\,,
\end{equation}
where
\begin{equation} \label{K2_mu_nu}
\mu = 2 (N-1) C_2, \quad
\nu = N \left(\frac{N+2}{3}\, C_0 C_4 - \frac{N-2}{2}\, C_2^2\right).
\end{equation}
Therefore, all the functions $C_0$, $C_2$ and $C_4$ satisfy the same second-order equation
\begin{equation} \label{K2_DS'-eq_C}
\ddot C = \mu\, \dot{C} + \nu \, C.
\end{equation}
Hence, the phase trajectories of system (\ref{K2_DS}) are planar curves.
(This is also seen from formulas (\ref{K2_C0,1,2})).
Note that
dynamical systems with a similar property appeared in \cite{FerHuaZha2012} when studying three-dimensional equations admitting the so-called ``central quadric ansatz''.

In the particular case $N=1$ from (\ref{K2_mu_nu}) we obtain $\mu=0$
and $\nu = C_0 C_4 + \frac{1}{2}\, C_2^2$ in accordance with formulas (\ref{DSt}), (\ref{q}) written for $C_1 = C_3 = 0$.

Taking into account (\ref{K2_C0,1,2}), the coefficients $\mu$ and $\nu$ can be represented in the form
\begin{equation} \label{K2_eq_C_coeff}
\begin{gathered}
\mu = -\frac{2 (N-1)}{N}\,\frac{\dot{y}}{y}\,, 
\\[4pt]
\nu = \frac{1}{2 N y^2}\,\left((N+2)\alpha-(N-2)\dot{y}^2\right) =
\frac{1}{N} \left(\frac{2 \alpha}{y^2} - \beta\, \frac{N-2}{2}\, y^{\frac{2}{N} - 1}\right),
\end{gathered}
\end{equation}
where $y(t)$ is a solution to equation (\ref{K2_eq_y}).
For $\beta \neq 0$, the functions $1/y$ and $\dot{y}/y$ form a fundamental system of solutions to equation (\ref{K2_DS'-eq_C}), (\ref{K2_eq_C_coeff}).

\newpage

\section{Investigation of non-invariant solutions (\ref{sol-1})}
\label{Ap_noninv-sol-1}

In this section we present a detailed study of all
types of non-invariant solutions described by formula (\ref{sol-1}).

It was noted above that (\ref{sol-1}) is reduced by transformation
(\ref{equiv2}) to the form (\ref{sol-1_simple}), which can be rewritten as
\begin{equation} \label{Ap8-sol}
v = \dfrac{\delta}{4 P} \left(x^4
 - 2\, \delta \, \dot{P}\, x^2 - g_3 \right)
= \dfrac{\delta}{4 P}
\left[ \left(x^2 - \delta \, \dot{P}\right)^2
- 4 P^3\right].
\end{equation}
Taking into account dilation (\ref{dilation1}),
we assume 
that $g_3 = \pm 1$.

We will be interested in the values of the arguments for which condition
\begin{equation} \label{Ap8-v>0}
v(x,t) > 0
\end{equation}
is satisfied and, thus, function
(\ref{u v}), $u(x,t) = (v(x,t))^{- 3/4}$,
 is defined.

Consider two cases, $g_3<0$ and $g_3>0$, with subcases corresponding to the choice $\delta = -1$ or $\delta = 1$.
In each case, for clarity, we present the graphs of both functions $v(x,t)$ and $u(x,t)$.

To obtain the asymptotics of solutions, we use the known expansions:

\smallskip
$P(t) = \dfrac{1}{t^2} + \dfrac{g_3}{28}\, t^4 + o(t^4)$ as $t \rightarrow 0$
\ and

 $P(t) = \dfrac{1}{(t-\tau)^2} + \dfrac{g_3}{28}\, (t-\tau)^4 + o((t-\tau)^4)$ as $t \rightarrow \tau$ \
 ($\tau$ is the period of $P(t)$).

\noindent

\medskip %
\noindent
If $g_3 < 0$, then $P(t)$ has two zeros on $(0,\tau)$:
$t_{01} = \tau/3$ and $t_{02} = 2 \tau/3$,
and

$P(t) = (-1)^i (-g_3)^{1/2} (t-t_{0i}) + \dfrac{1}{2}\,(-g_3) (t-t_{0i})^4 + o((t-t_{0i})^4)$
as $t \rightarrow t_{0i}$, $i=1,\,2$.

\medskip
Note that
$\tau \approx 5,2999$ for $g_3 = -1$ and
$\tau \approx 3,0599$ for $g_3 = 1$.

\subsection*{I. Case $g_3 < 0$}

In accordance with Remark~1 from Section~\ref{lame_sol},
three intervals for $t$ should be considered:
$(0, \tau/3)$, $(\tau/3, 2\tau/3)$,
$(2 \tau/3, \tau)$.
However, some of these cases can be excluded,
since condition (\ref{Ap8-v>0}) is violated for them.

Namely:
for $\delta = - 1$, $t \in (2 \tau/3, \tau)$,
the inequalities
$P(t)>0$ and $\dot{P}(t)>0$  hold,
and from (\ref{Ap8-sol}) we obtain
$v = -\frac{1}{4 P}\,\left[(x^2 + \dot{P})^2 - 4 P^3\right] < 0$,
since
\begin{equation*}
  (x^2 + \dot{P})^2 - 4 P^3 \geq \dot{P}^2 - 4 P^3
  = - g_3 > 0\,;
\end{equation*}
and for $\delta = 1$, $t \in (\tau/3, 2\tau/3)$, we have $P(t)<0$, and from (\ref{Ap8-sol})
it follows that $v < 0$.

Hence, it remains to consider four possibilities:
$\delta = - 1$, $t \in (0, \tau/3)$ or
$t \in (\tau/3, 2\tau/3)$,
\linebreak and
$\delta = 1$, $t \in (0, \tau/3)$ or
$t \in (2\tau/3, \tau)$.

\medskip
\noindent
\textbf{I.\,1.\,1.} \ $\delta = - 1$, \ $t \in (0, \tau/3)$.
\ Then $P(t)>0$, $\dot{P}(t)<0$
and from (\ref{Ap8-sol}) we obtain
\begin{equation*}
v = - \dfrac{1}{4 P}
\left(x^2 - \sigma_1\right)\left(x^2 - \sigma_2\right),
\end{equation*}
where
\begin{equation*}
\sigma_1(t) = - \dot{P} - 2 P^{3/2}, \ \
\sigma_2 = - \dot{P} + 2 P^{3/2}, \ \
\sigma_1 < \sigma_2.
\end{equation*}
Since $\sigma_1 \sigma_2 = \dot{P}^2 - 4 P^3 = - g_3 > 0$
and $\sigma_1 + \sigma_2 = - 2 \dot{P} > 0$,
then $\sigma_1$, $\sigma_2 > 0$.
Condition (\ref{Ap8-v>0}) is met if
$x \in (-\sigma_2^{1/2}, \, -\sigma_1^{1/2})\cup
(\sigma_1^{1/2}, \, \sigma_2^{1/2})$.

\begin{figure}[htbp]
\centering
\includegraphics[width=0.85\textwidth]
{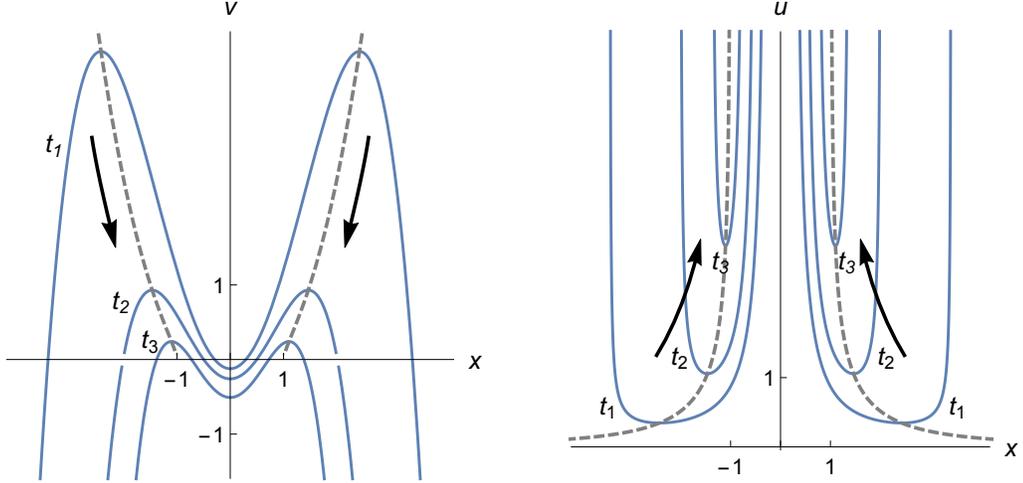}

\caption[]{Case I.1.1: \ $g_3 = -1$, \ $\delta = -1$, \ $t \in (0, \tau/3)$}

\label{fig_Ap8.I-1-1}
\end{figure}

Figure \ref{fig_Ap8.I-1-1} shows the graphs of solutions $v(x,t)$ and $u(x,t)$ for 
values of $t$ satisfying the condition
$0< t_1 < t_2 < t_3 < \tau/3$.
The dotted curves represent geometric places of local maxima of the function $v$ and, accordingly, local minima of the function
$u$.

\noindent

Since the functions $v$ and $u$ are even in $x$, consider the values $x > 0$.

When $t$ changes from $0$ to $\tau / 3$ the interval
$X(t) = (\sigma_1^{1/2}, \, \sigma_2^{1/2})$
monotonically contracts from $(0,+\infty)$ to point $1$.
As $t \rightarrow + 0$,  for any fixed $x \neq 0$,
we have $u(x,t) \rightarrow + 0$.
As $t \rightarrow  \tau/3 -0$, the following holds ($t_0 = \tau/3$):
\begin{equation*}
|X(t)| = \sigma_2^{1/2} - \sigma_1^{1/2}
\sim 2 (t_0-t)^{3/2} \rightarrow 0,
\ \
 \min_{X(t)} u(x,t) \sim (t_0-t)^{-3/2} \rightarrow +\infty,
\end{equation*}

Therefore, solution $u(x,t)$ describes a kind of blow-up regime: in a finite time the interval $X(t)$ contracts to a point, and the minimum of $u(x,t)$ on this interval tends to infinity.

\medskip

\noindent
\textbf{I.\,1.\,2.} \ $\delta = - 1$, \ $t \in (\tau/3, 2\tau/3)$.
Then $P(t)<0$ and condition (\ref{Ap8-v>0}) is met
for all $x \in (- \infty, + \infty)$.

\begin{figure}[htbp]
\centering
\includegraphics[width=0.85\textwidth]
{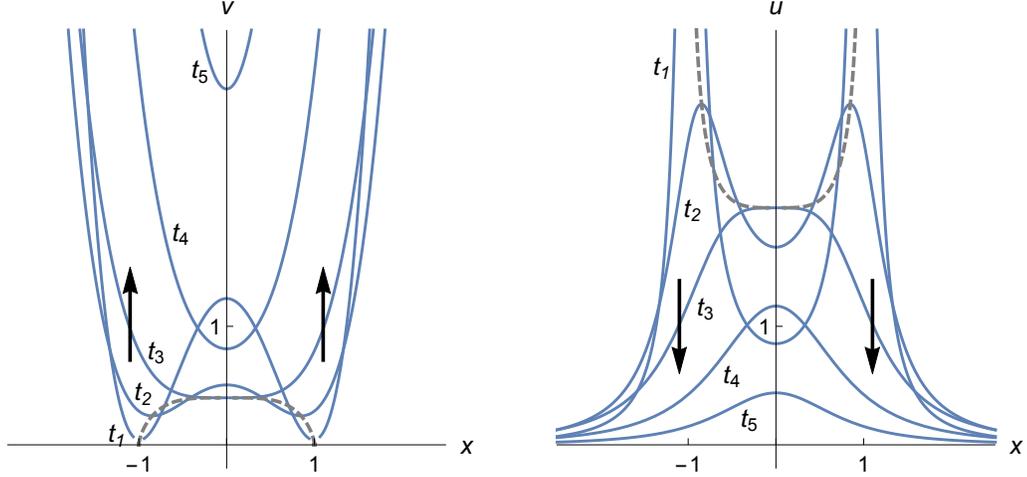}

\caption[]{Case I.1.2: \ $g_3 = -1$, \ $\delta = -1$, \ $t \in (\tau/3, 2\tau/3)$}

\label{fig_Ap8.I-1-2}
\end{figure}

Figure~\ref{fig_Ap8.I-1-2} shows the graphs of solutions $v(x,t)$ and $u(x,t)$ for
values of $t$ satisfying the condition
$\tau/3< t_1 < t_2 < t_3 =\tau/2 < t_4 < t_5 < 2 \tau / 3$.

At $t = \tau/2$, the solution profiles change qualitatively: for $t < \tau/2$ they have three extrema, and for $t \geq \tau/2$ -- only one extremum.
The dotted curves in Figure~\ref{fig_Ap8.I-1-2} are geometric places of local minima of the function $v$ and local maxima of the function $u$ at $t < \tau/2$.

\newpage
As $t \rightarrow  \tau / 3 + 0$, for any fixed $x$, we have
\begin{equation*}
u(x,t) \rightarrow
 \left\{
 \begin{array}{c}
 \begin{aligned}
       +\infty, \ \ \ &x = \pm 1,  \\
       +0, \ \ \ &x \neq \pm 1.
 \end{aligned}
 \end{array}
\right.
\end{equation*}
As $t \rightarrow  2 \tau / 3 - 0$, the function $u(x,t)$ uniformly tends to zero on $\mathbb{R}$.

\medskip
\noindent
\textbf{I.\,2.\,1.}
\ $\delta = 1$, \ $t \in (0, \tau/3)$.
\ Then $P(t)>0$, $\dot{P}(t)<0$ and
from (\ref{Ap8-sol}) we obtain
\begin{equation*}
v(x,t) > \frac{1}{4 P}
\left[ \dot{P}^2 - 4 P^3\right]
= - \frac{g_3}{4 P} > 0,
\end{equation*}
that is, condition (\ref{Ap8-v>0}) is met
for all $x \in (- \infty, + \infty)$.

\begin{figure}[htbp]
\centering
\includegraphics[width=0.85\textwidth]
{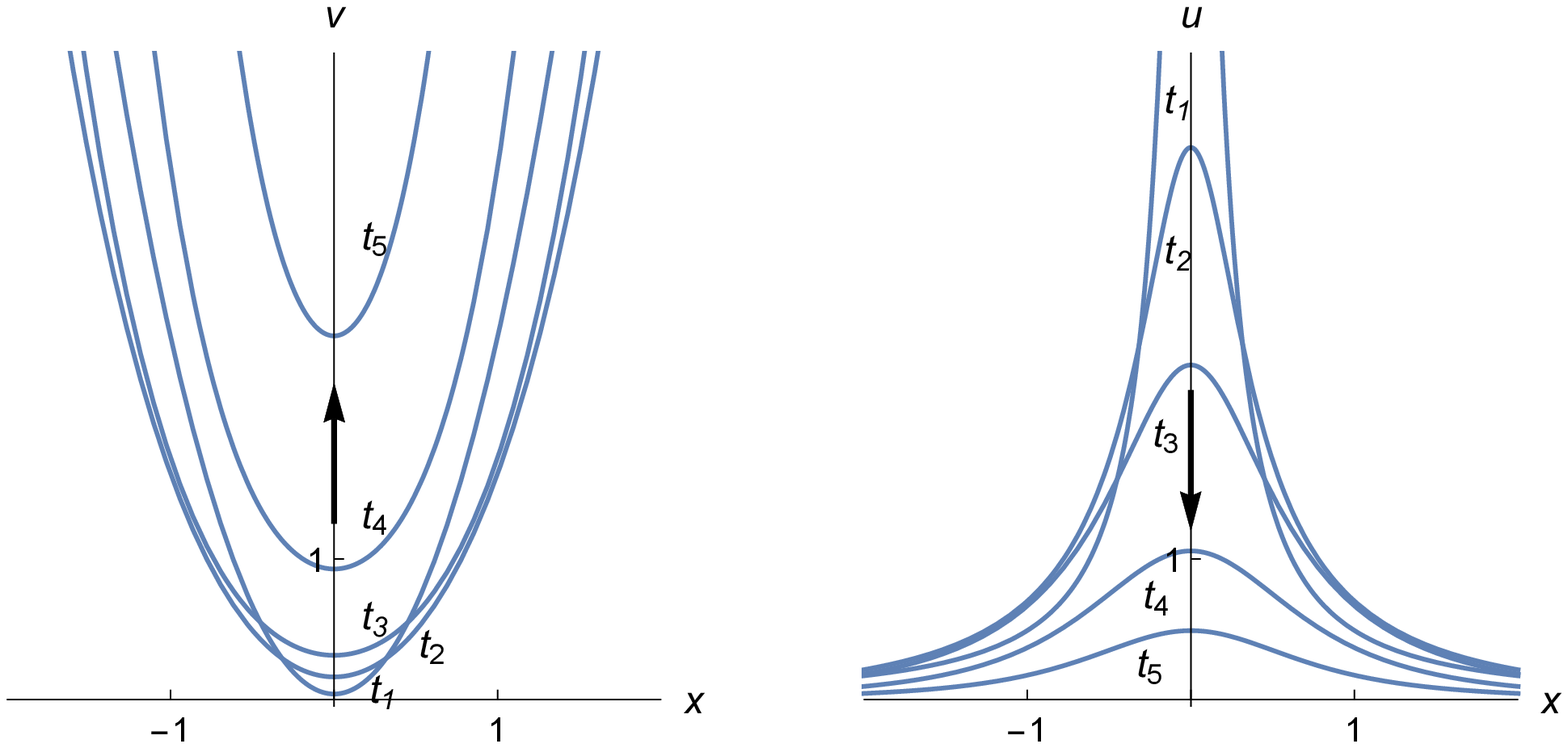}

\caption[]{
Case I.\,2.1\,: \ $g_3 = -1$, \
$\delta = 1$, \ $t \in (0, \tau/3)$
}
\label{fig_Ap8.I-2-1}
\end{figure}

Figure \ref{fig_Ap8.I-2-1} shows the graphs of solutions $v(x,t)$ and $u(x,t)$ for values of $t$ satisfying the condition
$0 < t_1 < t_2 < \dots < t_5 < \tau / 3$.

As $t \rightarrow + 0$, for any fixed $x$ we have:
\begin{equation*}
u(x,t) \rightarrow
 \left\{
 \begin{array}{c}
 \begin{aligned}
       +\infty, \ \ \ &x =0,  \\
       +0, \ \ \ &x \neq 0.
 \end{aligned}
 \end{array}
\right.
\end{equation*}
As $t \rightarrow  \tau / 3 - 0$, the function $u(x,t)$ uniformly tends to zero on $\mathbb{R}$.

\medskip
\smallskip
\noindent
\textbf{I.\,2.\,2.}
\ $\delta = 1$, \ $t \in (2\tau/3, \tau)$.
Then $P(t)>0$, $\dot{P}(t)>0$ and
from (\ref{Ap8-sol}) we obtain
\begin{equation*}
v = \dfrac{1}{4 P}
\left(x^2 - \sigma_1\right)\left(x^2 - \sigma_2\right),
\end{equation*}
where
\begin{equation*}
\sigma_1(t) = \dot{P} - 2P^{3/2}, \ \
\sigma_2(t) = \dot{P} + 2 P^{3/2}
\end{equation*}
and $0 < \sigma_1 < \sigma_2$ (as in case I.\,1.\,1).

Condition (\ref{Ap8-v>0}) is  met if
$x \in (-\infty,\,-\sigma_2^{1/2})\cup
(-\sigma_1^{1/2}, \, \sigma_1^{1/2})\cup
(\sigma_2^{1/2}, \, +\infty)$.

\begin{figure}[htbp]
\centering
\includegraphics[width=0.85\textwidth]
{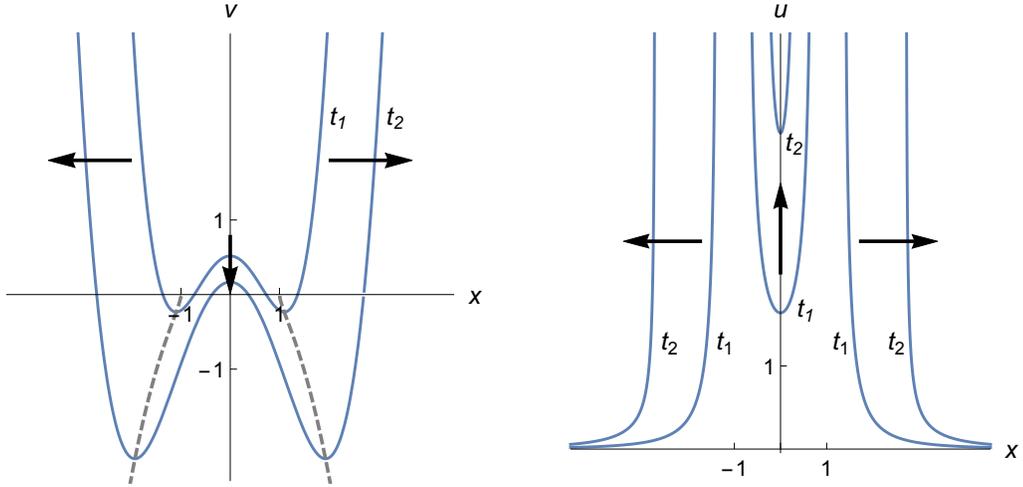}

\caption[]{Case I.\,2.2\,: \ $g_3 = -1$, \ $\delta = 1$, \ $t \in (2 \tau/3, \tau)$}

\label{fig_Ap8.I-2-2}
\end{figure}

Figure \ref{fig_Ap8.I-2-2} shows the graphs of solutions $v(x,t)$ and $u(x,t)$ for
values of $t$ satisfying the condition
$2\tau/3 < t_1 < t_2 < \tau$.
The dotted curve in the left figure is
a geometric place of local minima
of the function $v$.

When $t$ changes from $2 \tau / 3$ to $\tau$, $\sigma_1(t)$ decreases monotonically from 1 to 0, and $\sigma_2(t)$ increases from 1 to $+\infty$.
Thus, the interval $X(t) = (-\sigma_1^{1/2}, \, \sigma_1^{1/2})$
contracts from $(-1,1)$ to point $0$.
As $t \rightarrow  \tau -0$, we have
$\sigma_1^{1/2} \sim \frac{1}{2}\,(\tau-t)^{3/2} \rightarrow 0,
\
\sigma_2^{1/2} \sim  2 (\tau-t)^{-3/2} \rightarrow +\infty,
$
and
\begin{equation*}
 \min_{X(t)} u(x,t) \sim 2^{3/2} (t_0-t)^{-3/2} \rightarrow +\infty,
\end{equation*}

Solution $u(x,t)$ describes a blow-up regime similar to I.1.1:
in a finite time interval $X(t)$ contracts to a point, and the minimum of $u(x,t)$ on this interval tends to infinity.

\newpage

\subsection*{II. Case $g_3 > 0$}

In this case, the function $P(t)$ is defined and positive
on the interval $(0, \tau)$.
From (\ref{Ap8-sol}) we obtain
\begin{equation*}
v = \dfrac{\delta}{4 P}
\left[ \left(x^2 - \delta \, \dot{P}\right)^2
- 4 P^3\right]
= \dfrac{\delta}{4 P}
\left(x^2 - \sigma_1\right)\left(x^2 - \sigma_2\right),
\end{equation*}
where
\begin{equation*}
\sigma_1(t) = \delta \, \dot{P} - 2 P^{3/2}, \ \
\sigma_2(t) = \delta \, \dot{P} + 2 P^{3/2},
\end{equation*}
and, since $\sigma_1 \sigma_2 = \dot{P}^2 - 4 P^3 = - g_3 < 0$, we have
$\sigma_1 < 0 < \sigma_2$.

Condition (\ref{Ap8-v>0}) takes the form
$ \delta \left(x^2 - \sigma_2\right) > 0$,
that is, we need to consider two possibilities:
$\delta = -1$, $|x| < \sigma_2^{1/2}$ and
$\delta = 1$, $|x| > \sigma_2^{1/2}$.

\begin{figure}[htbp]
\centering
\includegraphics[width=0.85\textwidth]
{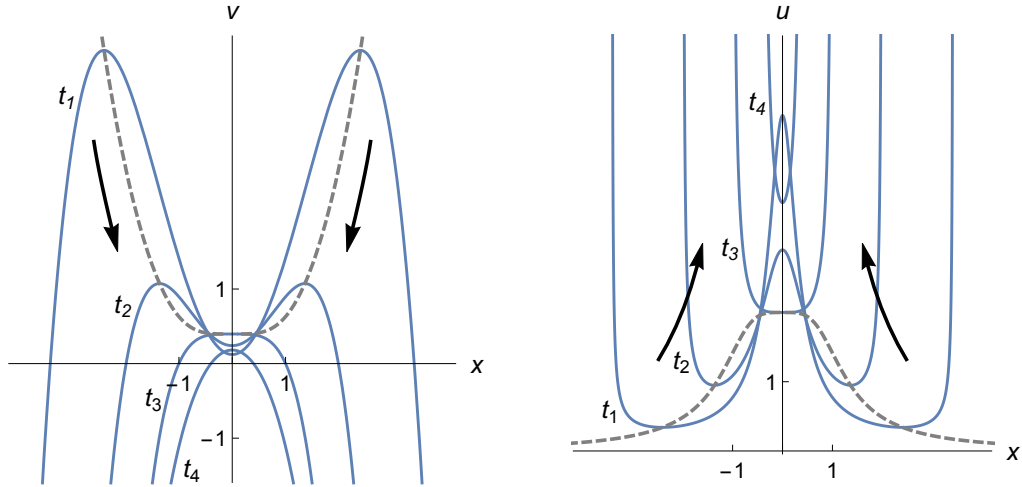}

\caption[]{Case II.\,1\,: \ $g_3 = 1$, \ $\delta = -1$, \ $t \in (0, \tau)$}

\label{fig_Ap8.II-1}
\end{figure}

\medskip
\noindent
\textbf{II.\,1.} \ $\delta = - 1$, \ $x \in (-\sigma_2^{1/2},\,\sigma_2^{1/2})$.
Figure \ref{fig_Ap8.II-1} shows the graphs of solutions $v(x,t)$ and $u(x,t)$ for 
values of $t$ satisfying the condition
$0< t_1 < t_2 < t_3 =\frac{\tau}{2}\, < t_4 < \tau$\,.

At $t = \tau/2$, the solution profiles change qualitatively: for $t < \tau/2$, they have three extrema, and for $t \geq \tau/2$ -- only one extremum.
The dotted curves in Figure~\ref{fig_Ap8.II-1} are geometric places of local maxima of the function $v$ and local minima of the function $u$ at $t < \tau/2$.

When $t$ changes from $0$ to $\tau$, the interval
$X(t) = (-\, \sigma_2^{1/2}, \, \sigma_2^{1/2})$
monotonically contracts from $(-\infty,+\infty)$ to point $0$.
As $t \rightarrow  + 0$, we have
$u(x,t) \rightarrow + 0$ for any fixed $x \neq 0$,
and $u(0,t) \rightarrow + \infty$.
As $t \rightarrow  \tau -0$, we obtain:

\begin{equation*}
|X(t)| = 2 \sigma_2^{1/2} \sim (\tau-t)^{3/2} \rightarrow 0,
\ \
 \min_{X(t)} u(x,t) \sim 2^{3/2} (\tau-t)^{-3/2} \rightarrow +\infty.
\end{equation*}

Solution $u(x,t)$ describes a blow-up regime similar to I.2.2 (and I.1.1):
in a finite time interval $X(t)$ contracts to a point, and the minimum of $u(x,t)$ on this interval tends to infinity.

\begin{figure}[htbp]
\centering
\includegraphics[width=0.85\textwidth]
{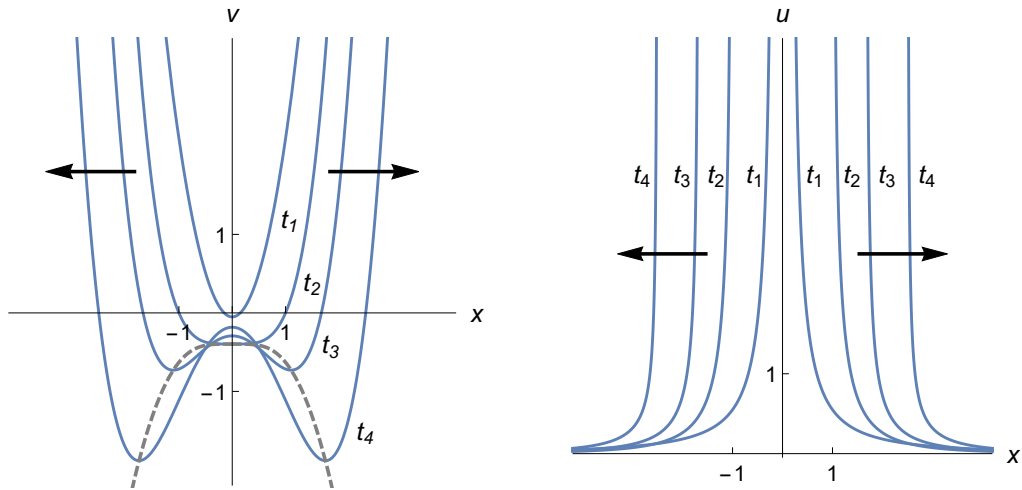}

\caption[]{Case II.\,2\,: \ $g_3 = 1$, \
$\delta = 1$, \ $t \in (0, \tau)$
}
\label{fig_Ap8.II-2}
\end{figure}

\medskip
\noindent
\textbf{II.\,2.} \ $\delta = 1$, \
$x \in (-\infty,\,-\sigma_2^{1/2})\cup (\sigma_2^{1/2},\,+\infty)$.
Figure \ref{fig_Ap8.II-2} shows the graphs of solutions $v(x,t)$ and $u(x,t)$ for
values of $t$ satisfying the condition
$0< t_1 < t_2 =\frac{\tau}{2}\, < t_3 < t_4 < \tau$\,.
At $t = \tau/2$, the profiles of solution $v(x,t)$ change qualitatively:
for $t \leq \tau/2$ they have one extremum, and for $t > \tau/2$ -- three extrema.

The function $\sigma_2(t) = \dot{P} + 2 P^{3/2}$ increases monotonically on the interval $(0, \tau)$, and $\sigma_2(+0) = +0$, $\sigma_2(\tau-0) = +\infty$.
When $t$ changes from $0$ to $\tau$, the set
$(-\infty,\,-\sigma_2^{1/2})\cup (\sigma_2^{1/2},\,+\infty)$
monotonically ``contracts'' from $(-\infty,0)\cup(0,+\infty)$ to the
empty set.
As $t \rightarrow + 0$, we have $u(x,t) \rightarrow + 0$ for any fixed $x \neq 0$.


\medskip

Summarizing, we note that in all the cases considered in this section,  solutions exist for a finite time and, at
any fixed $t$, are defined for $x \in \widetilde{X}(t)$, where $\widetilde{X}(t)$ is the union of some intervals in $\mathbb{R}$.

In cases I.1.1  and II.1,
peculiar blow-up regimes are realized: as $t$ increases, the set $\widetilde{X}(t)$  contracts to separate points, while the minimum of the solution $u(x,t)$ on $\widetilde{X}(t)$ tends to $+\infty$.
Solution I.2.2 can be interpreted similarly.
In cases  I.1.2 and I.2.1, for any admissible $t$, the set $\widetilde{X}(t)$ coincides with $\mathbb{R}$, and $u(x,t)$ tends to $+0$ uniformly,
fading out in a finite time
(although equation (\ref{heat}) has no zero solution).

Despite their peculiarity, the solutions obtained may be of interest to specialists studying blow-up in nonlinear evolution equations.



\section{King's solutions to equation (\ref{heat-m}) with
$N=1$, $m = - \frac{3}{2}$}
\label{King1993}

The next interesting example from \cite{King1993} is related to the equation
\begin{equation} \label{K_heat}
u_t = (u^{- 3/2}\, u_x)_x, 
\end{equation}
which, by substituting $u = v^{- 2/3}$, is reduced to the quadratic form
\begin{equation} \label{K_polyn4_eq}
v_t = v v_{2} - \frac{2}{3}\,v_1^2\,.
\end{equation}
The operator $F[v] = v v_{2} - \frac{2}{3}\,v_1^2$
possesses a four-dimensional invariant subspace
$W_4 = \mathcal{L}\{1,x,x^2,x^3\}$\,.
As a consequence, equation (\ref{K_polyn4_eq}) has
polynomial solutions 
\begin{equation} \label{K_polyn4_sol}
v = C_0(t) + C_1(t) x +  \frac{1}{2}\,C_2(t) x^2 + \frac{1}{6}\,C_3(t) x^3
\end{equation}
with coefficients satisfying the system
\begin{equation} \label{K_polyn4_DS}
\begin{aligned}
\dot C_0 &= C_0 \, C_2 - \frac{2}{3} \, C_1^2,
\
&\dot C_1 &= C_0 \, C_3 - \frac{1}{3} \, C_1 C_2,
\\
\dot C_2 &= \frac{2}{3} \, C_1 C_3 - \frac{1}{3} \, C_2^2, 
\
&\dot C_3 &= 0.
\end{aligned}
\end{equation}

In \cite{King1993}, non-invariant solutions
to equation (\ref{K_polyn4_eq}) on $W_4$ are
constructed in quadratures.
We will obtain an explicit representation for these solutions and justify their non-invariance.
Note that the ``method of differentiation''
applied above to 
(\ref{Gal_polyn5_DS}) does not work in
the case of system (\ref{K_polyn4_DS}).
Following \cite{King1993},
we will solve this system
by reducing its order using symmetries.

\subsection{All solutions of system (\ref{K_polyn4_DS}) and corresponding solutions (\ref{K_polyn4_sol})  of equation (\ref{K_polyn4_eq})}
\label{K_solutions}

Equation (\ref{K_polyn4_eq}) possesses a four-dimensional algebra of point symmetries with the basis
 \begin{equation} \label{K_polyn4_X}
X_1 = \frac{\partial}{\partial t}\,, \ \
X_2 = \frac{\partial}{\partial x}\,, \ \
X_3 = x \frac{\partial}{\partial x} + 2 v \frac{\partial}{\partial v}\,,
\ \ 
X_4 = t \frac{\partial}{\partial t} - v \frac{\partial}{\partial v}\,.
 \end{equation}
Algebra (\ref{K_polyn4_X}) induces the symmetry algebra $\{Y_1,\dots,Y_4\}$ of system (\ref{K_polyn4_DS}).
To reduce the order of system (\ref{K_polyn4_DS}), one can use the operator
\begin{equation*}
Y_2 = C_1 \frac{\partial}{\partial C_0} + C_2 \frac{\partial}{\partial C_1} + C_3 \frac{\partial}{\partial C_2}\,,
\end{equation*}
corresponding to $X_2$.
Its invariants (independent if $C_3 \neq 0$) are:
\begin{equation} \label{K_polyn4_IC}
I_1 = \frac{1}{9} \left( C_2^2 - 2 C_1 C_3 \right),
\
I_2 = - \frac{2}{9} \left( \frac{1}{3}\, C_2^3 - C_1 C_2 C_3 + C_0 C_3^2\right),
\
I_3 = C_3\,.
\end{equation}
On the solutions of system (\ref{K_polyn4_DS}), we obtain
\begin{equation} \label{K_polyn4_I12}
\dot{I}_1 = I_2, \ \	\dot{I}_2 = 6\,I_1^2,
\end{equation}
and, therefore, $I_1$ satisfies the equation
\begin{equation} \label{K_polyn4_DS-I1}
\dot{I}_1^2 = 4\,I_1^3 - g_3, \ \ g_3 = const.
\end{equation}
The third equation of system (\ref{K_polyn4_DS}) takes the form
\begin{equation} \label{K_polyn4_DS-I}
\dot C_2 = - 3\, I_1,
\end{equation}
while its fourth equation yields $C_3 = const$.

\smallskip
Consider two cases: $C_3 = 0$ and $C_3 = const \neq 0$.
(We represent solutions up to translations, dilations and reflections,
see Sections~\ref{lame_sol} and \ref{inv_sol}.)

\smallskip

1) For $C_3 = 0$, integrating system (\ref{K_polyn4_DS}),
we find the following solutions (\ref{K_polyn4_sol}):
\begin{equation} \label{K_inv-sol}
v = 1\,, \ \
v = \dfrac{3}{2}\,(x - t)\,, \  \
v = \dfrac{3}{2}\,\dfrac{x^2}{t}
\ \ \text{and} \ \
v =  \dfrac{3}{2}\,\dfrac{x^2}{t} + \delta\,t^3, \ \delta = \pm 1\,.
\end{equation}

2) Let $C_3 = const \neq 0$.
By using dilation (\ref{equiv2}) and reflection (\ref{refl-1}), we
set $C_3 = 3$,
and from (\ref{K_polyn4_IC}) we obtain
\begin{equation} \label{K_polyn4_CI}
C_0 = \frac{1}{18}
\left( \frac{1}{3}\, C_2^3 - 9\,I_1 C_2 - 9\,I_2 \right),
\ \
C_1 = \frac{1}{6} \left(C_2^2 - 9 I_1\right).
\end{equation}

From (\ref{K_polyn4_DS-I1}) and (\ref{K_polyn4_I12})
we have three possibilities:
if $g_3 = 0$, then a) $I_1=0$, $I_2=0$  or  b) $I_1=t^{-2}$, $I_2=-2t^{-3}$\,;
\ if $g_3 \neq 0$, then c) $I_1 = P(t) = \wp(t;\,0,g_3)$, $I_2 = \dot{P}(t)$.

In cases a) and b), we find solutions:
\begin{equation} \label{K_inv-sol-2}
v = x^3,
\ \ 
v = \dfrac{3}{2}\, \dfrac{x^2}{t} + x^3\,.
\end{equation}

It is directly seen that all solutions (\ref{K_inv-sol}) and (\ref{K_inv-sol-2}) are invariant.

In case c),
equation (\ref{K_polyn4_DS-I}) yields
\begin{equation*}
C_2 = 3\, \bigl(Z + S\bigr), \ \ S = const,
\end{equation*}
were $Z(t) = \zeta(t;0,g_3)$ is the Weierstrass $\zeta$-function,
$\dot{Z}(t) = - P(t)$,
and from (\ref{K_polyn4_CI}) we obtain
\begin{equation*}
C_1 = \frac{3}{2}\,\bigl((Z + S)^2 - P\bigr), \ \
C_0 = \frac{1}{2}\,\bigl((Z + S)^3 - 3 P\, (Z + S) - \dot{P}\bigr).
\end{equation*}
The substitution into (\ref{K_polyn4_sol}) leads to the expression
\begin{equation*}
v = \frac{1}{2} \left(\left(x + Z + S\right)^3 -
3 P \left(x + Z + S\right) - \dot{P}\right),
\end{equation*}
which, by translation in $x$, is reduced to the form
\begin{equation} \label{K_noninv-sol}
v = \dfrac{1}{2} \left(\bigl(x + Z\bigr)^3 - 3\, P \bigl(x + Z\bigr) - \dot{P}\right)
\,.
\end{equation}
This is an explicit representation of King's non-invariant
solutions of equation (\ref{K_polyn4_eq}).

Now we will show that solutions (\ref{K_noninv-sol}) are not invariant.


\subsection{Condition for non-invariance of solutions (\ref{K_polyn4_sol}). \\
Non-invariance of solutions (\ref{K_noninv-sol})}
\label{K2_noninv}

Similarly to Section~\ref{InvNoninv}, the invariance condition for solution (\ref{K_polyn4_sol}) leads to a system of linear homogeneous equations for the coefficients of a linear combination of basis operators (\ref{K_polyn4_X}).
The determinant of the matrix of this system has the form
\begin{equation} \label{K_Delta}
\Delta =
\frac{1}{3}\,C_3 \Bigl[ - 9 C_0^2 C_3^2 + 18 C_0 C_1 C_2 C_3
- 6 C_0 C_3^3 - 8 C_1^3 C_3 + 3 C_1^2 C_2^2 \Bigr].
\end{equation}
It is verified that $\Delta$ is the first integral of system (\ref{K_polyn4_DS}).

The condition $\Delta \neq 0$ is sufficient for the non-invariance of solution
(\ref{K_polyn4_sol}).
In the case of (\ref{K_noninv-sol}), we have
\begin{equation*}
C_0 = \frac{1}{2}\,\bigl(Z^3 - 3 P Z - \dot{P}\bigr), \ \
C_1 = \frac{3}{2}\,\bigl(Z^2 - P\bigr), \ \
C_2 = 3\, Z, \ \ C_3 = 3.
\end{equation*}
Substitution in (\ref{K_Delta}) yields $\Delta = \frac{81}{4}\, g_3 \neq 0$, so solutions (\ref{K_noninv-sol}) are non-invariant.

For invariant solutions (\ref{K_inv-sol}) and (\ref{K_inv-sol-2}),
we have $\Delta = 0$.

\subsection{Remark on solutions (\ref{K_noninv-sol})}
\label{K2_remark_qubic}

The explicit formula allows one to study solutions (\ref{K_noninv-sol}) in the similar way as it was done in Section~\ref{Ap_noninv-sol-1} for solutions (\ref{sol-1}).
We do not present a detailed
study here, confining ourselves to a remark about the roots of the cubic polynomial on the right-hand side of (\ref{K_noninv-sol}).

Putting $y = x + Z(t)$, we rewrite this cubic polynomial in the form
\begin{equation} \label{K_cubic}
y^3 + p\, y + q, \ \ p = -3 P, \ \ q = - \dot{P}.
\end{equation}
Its discriminant is
\begin{equation*}
D = - 4 p^3 - 27 q^2 = 27 g_3,
\end{equation*}
see, for example, \cite{Kur1972}.
Consequently,
the number of real roots of polynomial (\ref{K_cubic}) is determined by the sign of the invariant $g_3$:
for $g_3<0$ ($D<0$), the polynomial has one real root,
and for $g_3>0$ ($D>0$), it has three real roots.

Solution (\ref{K_noninv-sol}) is defined for any $x \in \mathbb{R}$ and $t \in (0, \tau)$, where $\tau$ is the period of the function $P(t)$.
It vanishes for values $x = \sigma_i(t)$, $i = 1,\dots,k$, that correspond to real roots of polynomial (\ref{K_cubic}) ($k = 1$ for $g_3<0$ and $k = 3$ for $g_3>0$).
The corresponding solution $u(x,t) = (v(x,t))^{- 2/3}$ of equation (\ref{K_heat}) is defined for any $t \in (0, \tau)$ and for any $x \in \mathbb{R}$ except for the points $x = \sigma_i(t)$, at which it
tends to infinity.


\section{Final remarks}

In this paper, \textit{all} solutions to equation (\ref{heat}) that have the form (\ref{u v}), (\ref{Gal_polyn5_sol}) are found; in particular, explicit formulas for non-invariant solutions are obtained.
Explicit representations are also given for non-invariant solutions constructed in \cite{King1993} to equations (\ref{K2_heat_n_m}) and (\ref{K_heat}).


In connection with the results obtained, a number of questions arise,
for example, concerning the properties of the found solutions
or the properties of nonlinear 
systems describing dynamics
of solutions on invariant subspaces.
Recall, for instance, the amazing properties of system (\ref{Gal_polyn5_DS}) established above: despite the rather high order of this system, all its phase trajectories are planar curves,
moreover, the system itself is radically simplified on differentiation.
(Compare with \cite{FerSvr2007}, where ODEs linearizable on differentiation were discussed.)

In conclusion, we give two more examples related to equations (\ref{heat}) and (\ref{Gal_polyn5_eq}).

1)
Adding a source or sink of a special type to the right-hand side of equation (\ref{heat}),
yields the equation
\begin{equation} \label{source}
u_t = (u^{-4/3}\, u_x)_x + \frac{3}{4}\, \delta\, u^{-1/3}\,, \ \ \delta = \pm 1\,,
\end{equation}
which, by substitution (\ref{u v}), $u = v^{- 3/4}$,
is reduced to the quadratic form
\begin{equation} \label{Gal_W5_eq}
v_t = v v_{2} - \frac{3}{4}\,v_1^2 - \delta\,v^2
\equiv F[v].
\end{equation}
It was established in \cite{Gal} that the operator $F[v]$ possesses
a maximal (five-dimensional) \textit{non-polyno\-mi\-al} invariant subspace:
for $\delta = 1$ -- \textit{exponential}, $W_5^{+} = \mathcal{L}\{1,e^{x},e^{-x},e^{2x},e^{-2x}\}$,
and for $\delta = - 1$ -- \textit{trigonometric},
$W_5^{-} = \mathcal{L}\{1,\cos{x},\sin{x},\cos{2x},\sin{2x}\}$.

Equation (\ref{Gal_W5_eq}) is related to equation (\ref{Gal_polyn5_eq}) (written
for $t$, $\bar{x}$, $\bar{v}$)
by changes of variables:
$\bar{x} = \pm e^{x}, \ \bar{v} = e^{ 2 x} v  \ \text{for} \  \delta = 1$,
or
$\bar{x} = 2\, \tan{(x/2)}, \  \bar{v} = \cos^{-4}{(x/2)}\, v \  \text{for} \  \delta = - 1$.
Note that these transformations also connect the invariant subspaces
$W_5^{+}$ and $W_5^{-}$ with polynomial
subspace (\ref{polyn5}) and, thus, solutions to equations (\ref{Gal_W5_eq}) on these subspaces can be obtained from solutions of equation (\ref{Gal_polyn5_eq}) on subspace (\ref{polyn5})
found in Sections~\ref{lame_sol} and \ref{inv_sol}.

\smallskip
2)
Along with equation (\ref{Gal_polyn5_eq}), $v_{t} = F[v]$, one can consider the wave equation $v_{tt} = F[v]$ with the same operator $F$ in the right-hand side.
This equation also possesses five-dimensional polynomial invariant subspace (\ref{polyn5}), as well as five-dimensional symmetry algebra.
In this case, it is also possible to consider solutions of the form (\ref{Gal_polyn5_sol}), but the order of the corresponding dynamical system
will double.

Similarly, one can consider the hyperbolic analogues of equations (\ref{K2_quadr_n_m}) and (\ref{K_polyn4_eq}).

\section{Acknowledgements}

The author thanks E.V.~Ferapontov, V.A.~Galaktionov and S.A.~Gutnik for clarifying discussions and useful remarks.



\begin{thebibliography}{99}
\addcontentsline{toc}{section}{References}

\bibitem{Akh1990}  N.I.~Akhiezer,
   \textit{Elements of the theory of elliptic functions}, Translated from the second
   Russian edition by H. H. McFaden. Translations of Mathematical Monographs, 79.
   American Mathematical Society, Providence, RI (1990). 

\bibitem{Bar1952} G.I.~Barenblatt, On some unsteady motions of a liquid and a gas in a porous medium, {\em Prikl. Mat. Mekh.} {\bf 16} (1952) 67--78.

\bibitem{BarZel1957} G.I.~Barenblatt and Ya.B.~Zel'dovich, On dipole-type solutions in problems of non\-stationary filtration of gas under polytropic regime, {\em Prikl. Mat. Mekh.} {\bf 21} (1957) 718--720.

\bibitem{Cra1956} J.~Crank, {\em The Mathematics of Diffusion}\,, Oxford (1956).

\bibitem{IbrHandbook1994} {\em  CRC Handbook of Lie Group Analysis of Differential Equations}, Vol.~{\bf 1}:~{\em Symmetries, Exact Solutions and Conservation Laws},   Ibragimov, N.H., Ed., CRC Press, Boca Raton, FL,~1994.

\bibitem{FerHuaZha2012}  E.V.~Ferapontov, B.~Huard and A.~Zhang, On the central quadric ansatz: integrable models and Painlev\'{e} reductions,
    \textit{J. Phys. A: Math. Theor.} \textbf{45}, no.~19 (2012) 195204.

\bibitem{FerSvr2007}  E.V.~Ferapontov  and S.R.~Svirshchevskii, Ordinary differential equations which linearize on differentiation,
    \textit{J. Phys. A: Math. Theor.} \textbf{40}, no.~9 (2007) 2037.


\bibitem{Gal}  V.A.~Galaktionov, Invariant subspaces and new explicit solutions to evolution equations with quadratic
nonlinearities, {\em Proc. Roy. Soc. Edinburgh, Sect.~A} {\bf 125} (1995) 225--246;
report AM-91-11, School of Math., University of Bristol (1991).

\bibitem{GalSvr}  V.A.~Galaktionov and S.R.~Svirshchevskii, {\em Exact Solutions and Invariant Subspaces of  Nonlinear Partial Differential Equations in Mechanics and Physics},  Chapman and Hall/CRC, Boca Raton, Florida (2007).

\bibitem{GomFerNov2021}  B.~Gormley, E.V.~Ferapontov and V.S.~Novikov, On a class of integrable Hamiltonian eqns in 2+1 dimensions, \textit{Proc. Roy. Soc. A} \textbf{477}: 20210047, 17 pp.

\bibitem{Hil1989} J.M.~Hill, Similarity solutions for nonlinear diffusion -- a new integration procedure, {\em J. Eng. Math.} {\bf 23} (1989) 141--155.

\bibitem{King1990} J.R.~King, Exact similarity solutions to some nonlinear diffusion equations, {\em J. Phys. A} {\bf 23} (1990) 3681--3697.

\bibitem{King1993} J.R.~King, Exact polynomial solutions to some nonlinear diffusion equations, {\em Physica D} {\bf 64} (1993) 35--65.

\bibitem{Kur1972} A.G.~Kurosh, {\em Higher algebra}, 
    Mir Publishers, Moscow (1972).

\bibitem{Olv1989eng} P.J.~Olver, \textit{Applications of Lie Groups to Differential Equations}, Second Edition, Graduate Text in Mathematics, vol. 107, Springer-Verlag, NY (1993). 

\bibitem{Ovs1959eng} L.V.~Ovsyannikov, Group properties of a nonlinear heat equation,
    {\em Dokl. Akad. Nauk SSSR} {\bf 125} (1959) 492--495.

\bibitem{Ovs1978eng} L.V.~Ovsyannikov, \textit{Group Analysis of Differential Equations,} Academic Press, NY (1982). 

\bibitem{Svr} S.R.~Svirshchevskii, Lie-B\"acklund symmetries of linear ODEs and generalized separation of variables in nonlinear equations, {\em Phys. Lett. A} {\bf 199} (1995) 344--348.

\bibitem{ZelKom1950} Ya.B.~Zel'dovich and A.S.~Kompaneets, On the theory of a heat propagation with heat conductivity depending on the temperature, {\em  Collection in Honor of the 70th Birthday of Academician of A.F.~Ioffe}, Izd. Akad. Nauk SSSR, Moscow (1950) 61--71.


\end{thebibliography}
\end{document}